\documentstyle[preprint,prd,eqsecnum,aps,epsf]{revtex}

\newcounter{num}
\newcommand{\sq}[2]{\tilde{#1}_{#2}}
\newcommand{\be}{\begin{equation}}
\newcommand{\ene}{\end{equation}}
\newcommand{\bea}{\begin{eqnarray}}
\newcommand{\enea}{\end{eqnarray}}
\newcommand{\ee}{e^+e^-}
\newcommand{\dz}[2]{\delta Z_{#1}^{#2}}
\newcommand{\ff}{\frac{1}{2}(Z_{\tilde{t}}^{1i})^2-\frac{2}{3}s_W^2}
\newcommand{\coupl}[1]{{\tilde {t_i}}^\ast\stackrel{\leftrightarrow}{\partial^\mu}\tilde{t_i} #1_\mu}
\newcommand{\dtheta}[1]{\frac{\delta\cos\theta_W}{\cos\theta_W}\ #1 
\frac{\delta\sin\theta_W}{\sin\theta_W}}
\newcommand{\mass}[2]{m_{{\tilde{#1}}_#2}^2}
\newcommand {\dmass}{\frac{\delta M_Z^2}{M_Z^2}-\frac{\delta M_W^2}{
M_W^2}}

\begin{document}
\vspace{1.cm}

\begin{center}
{\Large\bf Large Yukawa coupling corrections to scalar quark pair production 
in $e^+e^-$ annihilation}
\end{center}
\vspace{0.5cm}

\begin{center}
{ Xiao-Jun Bi, Yuan-Ben Dai and Xiao-Yuan Qi}\\\vspace{3mm}
{\it Institute of Theoretical Physics,
 Academia Sinica, \\ P.O.Box 2735, Beijing 100080, P. R. China }
\end{center}
\vspace{1.5cm}

\begin{abstract}
We calculate the large Yukawa coupling corrections to the top and bottom
scalar quark pair production in $\ee$ annihilation within the
Minimal Supersymmetric Standard Model. We include the vertex corrections and the corrections to the gauge boson propagator enhanced by large masses. We find the total corrections are quite
significant. In some regions of the parameter space the corrections are larger than 10\%.

\end{abstract}

\section {INTRODUCTION}

Supersymmetry(SUSY) is one of the most attractive extensions of the
Standard Model(SM). It provides an elegant way to stabilize the huge
hierarchy between the electroweak and the GUT scales against radiative
corrections\cite{hierar}. Moreover, supersymmetric models offer
a natural solution to the Dark Matter problem\cite{darkm} and allow for
a consistent unification of the all known gauge coupling constants
in contrast to the SM\cite{uni}.
Due to the theoretical appealing of SUSY, the search for
supersymmetric particles is one of the main issues in the experimental
programs at the CERN $\ee$ collider LEP2 and Fermilab Tevatron
\cite{exp}. It will play an even more important role at the future
Large Hadron Collider (LHC)\cite{lhc} and the Next $\ee$ Linear Collider\cite{nlc}.

Although the colored supersymmetric particles, squarks and gluinos,  
can be searched for most efficiently at hadron colliders, for a precise
determination of the underlying SUSY parameters lepton colliders will 
be necessary. 
%Furthermore, because of the large background, 
%the third generation squarks
%may be difficult to find at the LHC while they can be easily detected
%in future $\ee$ colliders\cite{ee}. 
For the experimental search 
it is useful to predict the production rates 
of these particles precisely incorporating radiative corrections.
Up to now, many works have been devoted to the QCD corrections to
various sparticle production rates.
QCD corrections to colored sparticle (except stop) production 
at hadron colliders
were discussed in detail by W. Beenakker {\it et al.}~\cite{ccsq}.
The corresponding corrections to the top squark production were given
in another paper\cite{ccst}. The QCD and SUSY-QCD corrections to non-colored
sparticle production at hadron colliders were given in~\cite{cnsq}
and those to colored sparticle production at 
$\ee$ colliders were given in~\cite{eesq}.

In this paper, we consider the electroweak corrections to the 
third generation diagonal squark pair production in $\ee$ annihilation,
$\ee \rightarrow \sq{t}{i}{\bar{\sq{t}{i}}}, \sq{b}{i}{\bar{\tilde{b}}_{i}}$,
due to large Yukawa couplings.
% The corrections arise from two 
%effects, that is, the direct couplings between the stop or sbottom
%and the Higgs bosons or higgsinos and gauge bosons masses corrections
%due to virtual heavy particles loops. 
Our framework is the Minimal Supersymmetric Standard Model(MSSM)\cite{mssm}.
As is well known that there are five physical Higgs bosons in the MSSM, two CP-even neutral Higgs bosons, one CP-odd neutral Higgs boson and a pair of charged Higgs bosons. Their supersymmetric partners, higgsinos, are components of
two charginos and four neutralinos in the MSSM.
The top and bottom squarks have Yukawa couplings with these Higgs bosons and 
higgsinos, which are proportional to $m_t\cot\beta$ or $m_b\tan\beta$, where
% squarks 
%may have large Yukawa coupling with the Higgs bosons and higgsino for large
$\tan\beta=v_2/v_1$ and $v_1$, $v_2$ are the vacuum expectation
values of the two Higgs doublets. These interaction terms are large in the region of small or large $\tan\beta$ and they can even be leading electroweak corrections for $\tan\beta\sim 1$ or $\tan\beta\sim {m_t/m_b}$. On the other hand, the internal gauge bosons may also
have large corrections enhanced by large masses due to virtual heavy particle loops such as the top or stop loops. For consistency, we also include such corrections in our calculations. 
We calculate in the 't Hooft-Feynman gauge.
We find the total corrections are quite large in some regions of the MSSM parameter space allowed by present experiments, 
which can be larger than the SUSY-QCD corrections to the same process
due to gluino exchanges\cite{eesq}. 
%Throughout the paper we shall retain only terms proportional to $m_t$, $m_b\tan\beta$ or the large masses of squarks, Higgs bosons and higgsinos.

This paper is organized as follows. In Sec. II we present the renormalization
scheme adopted in our calculation.
Some analytic results are given in Sec. III and the numerical results 
are discussed in Sec IV.
Finally we summarize the conclusion in Sec V. The relevant pieces of the Lagrangian are presented
in Appendix A and some analytic expressions are collected in Appendix B.

\section {RENORMALIZATION SCHEME}
In this section we briefly discuss the renormalization scheme 
adopted in our calculations. To calculate 
the electroweak corrections to the process $\ee\rightarrow\sq{t}{i}
{\bar{\sq{t}{i}}}(\sq{b}{i}{\bar{\tilde{b}}}_{i})$ at one loop level,
we do not need to consider the renormalization of the 
Higgs sector after imposing the vanishing of the tadpole terms. (However, we adopt an approximate Higgs mass 
formula including radiative corrections.)
Thus, the renormalization scheme
is focused on the gauge sector. It differs only slightly from that given
by M. B\"ohm\cite{bohm}. Another complexity arises from the renormalization
of the squark mixing angle.
%Since we concern the radiative corrections
%of order ${\cal O}(\alpha m_t^2/M_W^2)$ given by exchanging virtual
%Higgs boson or virtual top quark loop(see Fig. 3,4,5)
%we will see that many correction terms are
%zero and our calculation are considerably simplified.

\subsection {Gauge boson renormalization}

The diagonal production of squark pairs proceeds through S-channal
photon and $Z$ boson exchange at tree level(see Fig. 1). 
The longitudinal part of $Z$ boson does not give any contribution to 
the process. In the MSSM the
photon and $Z$ boson may mix with the CP-odd neutral Higgs boson $A^0$
and the neutral Goldstone boson $G^0$ at one loop level\cite{oneloop}.
However, for diagonal production of squark pairs, 
such mixing does not give any contribution either.
So only the renormalization of the transverse part of  
the gauge bosons is needed.

To respect gauge symmetry explicitly, each gauge multiplet is associated with one 
renormalization constant\cite{bohm}:
\bea
W_\mu^a\rightarrow (Z_2^W)^{1/2}\, W_\mu^a\ , \qquad
B_\mu\rightarrow (Z_2^B)^{1/2}\, B_\mu\ ,\nonumber \\
g_2\rightarrow Z_1^W(Z_2^W)^{-3/2}\, g_2\ , \qquad
g_1\rightarrow Z_1^B(Z_2^B)^{-3/2}\, g_1 \ \ .
\enea
The Weinberg angle $\theta_W$ is defined by the on-shell condition
$\cos\theta_W=\frac{M_W}{M_Z}$,
where $M_W$ and $M_Z$ are the masses of $W$ and $Z$ bosons.
Now we denote
\be
c_W=\cos\theta_W\ ,\qquad s_W=\sin\theta_W
\ene
as abbreviations throughout the paper and  
\bea
\delta Z_i^\gamma&=&s_W^2\delta Z_i^W+c_W^2\delta Z_i^B\,\ ,\qquad
\delta Z_i^Z=s_W^2\delta Z_i^B+c_W^2\delta Z_i^W\,\ ,\nonumber\\
\delta Z_i^{\gamma Z}&=&-c_Ws_W(\delta Z_i^W-\delta Z_i^B)
\enea
as the renormalization constants
 for the photon, $Z$ boson and $\gamma -Z$ mixing terms respectively.
Then we get 
\be
\label{gaz}
{ Z \choose A } \rightarrow \left( \begin{array}{cc}
1+\frac{1}{2}\delta Z_2^Z & -\delta Z_1^{\gamma Z}+\delta Z_2^{\gamma Z}\\
\delta Z_1^{\gamma Z}-2\delta Z_2^{\gamma Z}& 1+\frac{1}{2}\delta Z_2^\gamma \end{array}
\right)
{Z \choose A }
\ene
from which we can see the $\gamma-Z$ mixing term.

The renormalization constants $Z_{1,2}^W,\ Z_{1,2}^B$ are fixed by
the following on-shell conditions
\bea
\label{vvv}
\hat\Sigma_T^W(M_W^2)=\hat\Sigma_T^Z(M_Z^2)=\hat\Sigma_T^{\gamma Z}(0)=0\ ,\\
\hat\Gamma_\mu^{\gamma ee}(k^2=0,p\hspace{-2mm} \slash=q\hspace{-2mm} \slash=0)=ie\gamma_\mu\ ,\\
\label{ver}
\frac{1}{k^2}\hat\Sigma^\gamma(k^2)|_{k^2=0}=0
\enea
where the $\hat\Sigma_T$s represent the renormalized self-energies
and the $\hat\Gamma^{\gamma ee}$ represents the renormalized photon-electron vertex.
%After imposing the vanishing of the tadpole terms, 
From $M_W=g_2/2\sqrt{v_1^2+v_2^2}$ and 
$M_Z=\frac{1}{2}\sqrt{g_1^2+g_2^2}\sqrt{v_1^2+v_2^2}$ we get
%omitting the longitudinal part of 
%Z boson and the mixing with Higgs boson and only considering the
%corrections due to large Yukawa couplings, 
\bea
\label{de}
{{\delta M}_W^2 \over M_W^2}-{{\delta M}_Z^2 \over M_Z^2}=s_W^2\left[(2\delta Z_1^W-3\delta Z_2^W)-(2\delta Z_1^B-3\delta Z_2^B)\right].
%{{\delta M}_W^2\over M_W^2}-{{\delta M}_Z^2\over M_Z^2}&=&\delta g_2^2 (v_1^2+v_2^2)/4=M_W^2(2\dz{1}{W} -3\dz{2}{W} )\ ,\nonumber \\
%{\delta M}_Z^2&=&M_Z^2[(2 \dz{1}{W} -3\dz{2}{W} )c_W^2+(2\dz{1}{B} -3\dz{2}{B} )s_W^2]\ . 
\enea
Throughout this paper we shall keep only corrections proportional to a large mass $M>M_Z$. All terms independent of the large masses or depending on them only logarithmically will be ommitted. It is found that no terms proportional to large masses enter the expressions ${\Sigma^{\gamma}(k^2)\over k^2}|_{k^2=0}$ and $\Sigma^{\gamma Z}(0)$. The same is true for $\delta Z_1^{\gamma}$ determined from (2.6). Taking into account these facts we obtain from the renormalization conditions (2.5)-(2.7)
%Using (\ref{vvv})-(\ref{ver}) and neglecting terms which depend on the large masses only logarithmically, we find
\bea
\label{de2}
s_W^2\delta Z_2^W+c_W^2\delta Z_2^B&=&\dz{2}{\gamma}=-{\Sigma^\gamma(k^2)\over k^2}|_{k^2=0}=0\ , \nonumber \\
-\dz{1}{\gamma Z} +\dz{2}{\gamma Z} &=&\frac{\Sigma^{\gamma Z}(0)}{M_Z^2}=0\ ,\nonumber \\
\dz{1}{\gamma}&=&0\ .
\enea
The calculations here are similar to those in \cite{bohm}.
The four equations in (\ref{de}) and (\ref{de2}) completely determine all wave function renormalization constants as only four of them are independent. We get from these equations
\bea
\label{zz}
\dz{2}{Z} &=&\frac{c_W^2-s_W^2}{s_W^2}\left( \frac{{\delta M}_Z^2}{M_Z^2}-
\frac{{\delta M}_W^2}{M_W^2}\right), \nonumber \\
\label{gammaz}
\dz{2}{\gamma Z} &=& \frac{-c_W}{s_W}\left( \frac{{\delta M}_Z^2}{M_Z^2}-
\frac{{\delta M}_W^2}{M_W^2}\right) .
\enea

The self-energies of gauge bosons $\Sigma^\gamma$, $\Sigma^{\gamma Z}$
and $\Sigma_T^Z$(Fig. 2a) for $k^2\neq 0$ may contain terms proportional to large masses. However, it turns out that their contribution to renormalized gauge boson propagators can be neglected. As an example, let us look into the top-quark loop correction to the renormalized $Z$ boson propagator. This propagator can be written as
\bea
\label{reprop}
-ig_{\mu\nu}\left({1\over{k^2-M_Z^2}}-{1\over{k^2-M_Z^2}}(\Sigma_T^Z(k^2)-{\Sigma}_T^Z(M_Z^2)){1\over{k^2-M_Z^2}}-{\delta Z_2^Z \over k^2-M_Z^2}\right).
\enea 
%which do not contain terms proportional to $(\frac{m_t}{M_Z})^2$ 
The contribution of the top-quark loop to $\Sigma_T^Z(k^2)$ is written down in (B.20). Although $\Sigma_T^Z$ contains terms proportional to $m_t^2$, it can be checked that the combination $(\Sigma_T^Z(k^2)-{\Sigma}_T^Z(M_Z^2))\over{k^2-M_Z^2}$ is not  enhanced by $m_t^2$ and can be negelected compared to $\delta Z_2^Z$ in (\ref{reprop}) for all values of $k^2$ (See discussion following (B.21)). Thus, $Z$ and $\gamma-Z$ boson propagator can be written as
\be\label{pro2}\frac{-ig^{\mu\nu}}{k^2-M_Z^2}(1-\dz{2}{Z})\ene and
\be\label{pro}\frac{-ig^{\mu\nu}}{k^2-M_Z^2}\dz{2}{\gamma Z}\ene 
respectively.

The analytic expressions for the gauge boson mass corrections calculated from Fig. 3 are given in Appendix B. By using (B.15)-(B.19) we have checked that the divergences in individual terms of the expression (B.4) for ${\delta M_W^2 \over M_W^2}-{\delta M_W^2 \over M_Z^2}$ are cancelled out after omitting terms not enhanced by a large mass $M>M_Z$. Hence $\delta Z_2^Z$ and $\delta Z_2^{\gamma Z}$ obtained from (\ref{zz}) are finite.  This is also confirmed numerically. It should be noted that the full expression for $\delta Z_2^\gamma$ and $\delta Z_2^{\gamma Z}$ are of course divergent. The finite results obtained here are consequences of omitting divergent terms independent of large masses $M>M_Z$. After cancellation of divergences in the full expression such terms can not induce finite corrections proportional to a large mass.
\subsection{Renormalization of squark wave function}
There are two scalar partners $\sq{q}{L}$ and $\sq{q}{R}$ 
for every quark $q$ in SUSY theories. They mix and form 
two mass eigenstates $\sq{q}{1}$, $\sq{q}{2}$ 
which are related to the original fields by 
\be
{\sq{q}{L}\choose\sq{q}{R}}\ =\ Z_{\sq{q}{}}{\sq{q}{1}\choose\sq{q}{2}}
\ene
where \be
Z_{\sq{q}{}}\ =\ \left(\begin{array}{cc}
\cos\theta_{\sq{q}{}}&-\sin\theta_{\sq{q}{}}\\
\sin\theta_{\sq{q}{}}&\cos\theta_{\sq{q}{}}\end{array}\right)\ .\ene
%To be consistent, one should adopt a supersymmetric renormalization scheme
%which defines one common wave function renormalization constant for
%fermion and its supersymmetric partner sfermion. However, since the
%supersymmetry is explicitly broken
We will adopt a scheme in which 
both stops and sbottoms are defined on shell. We give the formulas
for stops here while those for sbottoms are similar.

The complexity of the squark wave function renormalization
is due to the fact that the two diagonalized states $\sq{q}{1}$ and 
$\sq{q}{2}$ mix again at one loop level(See Fig. 4).
Write the bare stop fields as
\be
\sq{t}{i}^0=\left( 1+\frac{1}{2}\dz{i}{\sq{t}{}} \right)
%\sq{t}{i}\rightarrow \left( 1+\frac{1}{2}\dz{i}{\sq{t}{}} \right)
\tilde{t}_i + 
\dz{ij}{\tilde{t}}\tilde{t}_j\, \ ,\quad j\neq i \ .
\ene
(We use $\delta Z^{\tilde q}_{i}$ and $\delta Z^{\tilde q}_{ij}$ to represent the wave function renormalization constants
and $Z_{\sq{q}{}}$ the mixing matrix.) 
The on-shell renormalization conditions require
that the mass parameters are the physical masses, the residues of
the squark propagators on shell are one and the mixing between on-shell squarks should be absent, {\it i.e.}
\bea
&&\hat\Sigma^{1i}(\mass{t}{1}) = 0\, \ ,\quad
\hat\Sigma^{2i}(\mass{t}{2})=0\,\ ,\quad i=1,2\ \ ,\nonumber\\
&&\frac{d}{dp^2}\hat\Sigma^{11}(p^2)|_{p^2=\mass{t}{1}}=0\,\ ,\quad
\frac{d}{dp^2}\hat\Sigma^{22}(p^2)|_{p^2=\mass{t}{2}}=0\,\ .
\enea
From the above equations, we get 
\bea
\delta m_{\sq{t}{i}}^2&=&\Sigma(m_{\sq{t}{i}}^2)\,\ ,\nonumber\\
\dz{i}{\sq{t}{}}&=&-\Sigma'(\mass{t}{i})\,\ ,\quad
\dz{ij}{\sq{t}{}}=\frac{\Sigma^{ji}(\mass{t}{j})}{\mass{t}{i}-\mass{t}{j}}
\, \ ,\enea
where $\Sigma'(p^2)$ is the derivative of $\Sigma (p^2)$ with respect to $p^2$.

The wave function renormalization constant matrix can be decomposed into
a symmetric and an antisymmetric part
\be
\label{angle}
{\sqrt {Z}}=\left(
	\begin{array}{cc}
	1+\frac{1}{2}\dz{1}{\sq{t}{}}&\frac{1}{2}(\dz{12}{\sq{t}{}}+\dz{21}{\sq{t}{}})\\
	\frac{1}{2}(\dz{12}{\sq{t}{}}+\dz{21}{\sq{t}{}})&1+\frac{1}{2}\dz{2}{\sq{t}{}}
	\end{array}\right)\cdot
\left(
	\begin{array}{cc}
	1&\frac{1}{2}(\dz{12}{\sq{t}{}}-\dz{21}{\sq{t}{}})\\
	-\frac{1}{2}(\dz{12}{\sq{t}{}}-\dz{21}{\sq{t}{}})&1
	\end{array}\right)\qquad,
\ene
where the off-diagonal elements of the symmetric part are ultraviolet finite and the antisymmetric part can be interpreted as a rotation matrix in the first order. Besides the wave function and gauge coupling constant renormalization defined above, an additional renormalization of the stop mixing angle $\theta_{\sq{t}{}}\rightarrow\theta_{\sq{t}{}}+\delta\theta_{\sq{t}{}}$ must be introduced to make the $Z\sq{t}{i}{\bar{\sq{t}{j}}}$ vertex part finite beyond the tree level. We choose $\delta\theta_{\sq{t}{}}$ such that this additional rotation just cancels the last factor in ({\ref{angle}}), that is,
\be \label{angle2} \delta\theta_{\tilde{t}}=\frac{\dz{12}{\tilde{t}}-\dz{21}{\tilde{t}}}
{2}\ .\ene This is the same scheme as used in \cite{sola}. 
It is found that with this choice of mixing angle renormalization the ultraviolet
divergence in the vertex graph is exactly cancelled.
%one must impose a condition related to the renormalized
%states $\sq{t}{L}$ and $\sq{t}{R}$. Write
%\be
%\sq{t}{L,R}\rightarrow\left( 1+\frac{1}{2}\dz{L,R}{\sq{t}{}} \right)
%\tilde{t}_{L,R} + 
%\dz{LR(RL)}{\tilde{t}}\tilde{t}_{R,L}\,\ ,
%\ene
%we demand a somewhat artificial condition requiring that $\dz{LR}{\tilde{t}}=
%\dz{RL}{\tilde{t}}$. Then we have
%\be\delta\theta_{\tilde{t}}=\frac{\dz{12}{\tilde{t}}-\dz{21}{\tilde{t}}}
%{2}\ .\ene This is the same result as given in \cite{sola}.

The analytic expressions for the self energies $\Sigma^{ij}$s
calculated from Fig. 4 are given in Appendix B.
\subsection{Renormalization of the gauge boson and squark vertex}
%Finally we discuss the renomalization of the gauge boson and the top squark
%interaction vertex.(That for sbottom vertex is similar.)
With the choice of (\ref{angle2}) the
% renormalized Lagrangian plus counter terms for the gauge boson and stop interaction vertex is given by
complete one-loop electroweak corrected Lagrangian for the gauge boson and stop interaction vertex is given by 
%The relevant vertices at one-loop level are
\bea
\label{gtt}
{\cal L}_{ \gamma\sq{t}{i}\bar{\sq{t}{i}} }&\ =\ &\left\{
-\frac{2}{3}ie\left( 1+\frac{\delta e}{e}+\dz{\sq{t}{i}}{}+\frac{1}{2}
\dz{2}{\gamma}\right)-\frac{ie}{s_Wc_W}\left(\ff\right)(-\dz{1}{\gamma Z}+
\dz{2}{\gamma Z})\right\}\coupl{A}\ ,\nonumber\\ \\
\label{ztt}
{\cal L}_{Z \sq{t}{i}\bar{\sq{t}{i}} }&\ =\ &
-\frac{ie}{s_Wc_W}\left[\ff\right]\left( 1+\frac{\delta e}{e}-\dtheta{-}+\dz{\sq{t}{i}}{}
+\frac{1}{2}\dz{2}{Z}\right)\coupl{Z}\nonumber \\
&&-\frac{ie}{s_Wc_W}\left(-\sin \theta_{\sq{t}{}}\cos \theta_{\sq{t}{}}\frac
{\dz{12}{\tilde{t}}+\dz{21}{\tilde{t}}}{2}\right)\coupl{Z}\nonumber\\
&&-\frac{2}{3}ie(-\dz{1}{\gamma Z}+
\dz{2}{\gamma Z})\coupl{Z}\ .
\enea
By using (\ref{de2}), the above two equations are reduced to  
\bea
{\cal L}_{ \gamma\sq{t}{i}\bar{\sq{t}{i}} }
&\ =\ &-\frac{2}{3}ie(1+\dz{\sq{t}{i}}{})\coupl{A}\ , \\
{\cal L}_{Z \sq{t}{i}\bar{\sq{t}{i}} }
&=&-\frac{ie}{s_Wc_W}\left[\ff\right](1+\dz{\sq{t}{i}}{})\coupl{Z}\nonumber\\
&&-\frac{ie}{s_Wc_W} 
\left(-\sin\theta_{ \sq{t}{} }\cos \theta_{ \sq{t}{} }
\frac{ \dz{12}{\tilde{t}}+\dz{21}{\tilde{t}} }{2}\right)\coupl{Z}\ .
\enea
The corresponding Lagrangian for the sbottoms is similar.
%The corrections to the gauge boson and bottom squark interaction vertex are similar.
%The renormalization of the vertex are simplified greatly
%for we are only concerned about the large Yukawa corrections.
%The Z boson renormalization constant $\dz{2}{Z}$ can be transferred to
%the vertex by replacing
%\be
%1+\dz{t}{i}\rightarrow 1+\dz{t}{i}-\dz{2}{Z}
%\ene
\section {ANALYTIC RESULTS}
\subsection{Vertex corrections}
All the quantum effects, including the vertex corrections and the corrections to the $Z$ gauge boson propagator but not the $\gamma -Z$ mixing contributions, can be written in concise forms by defining two effective coupling constants $D^\gamma$ and $D^Z$. 
Let
\bea
\label{vg}
\Gamma_{eff}^{\gamma \sq{t}{}\sq{t}{}}&=&-ie\left[
\frac{2}{3}(1+\dz{\sq{t}{i}}{})\ +\ \Lambda^\gamma\right]\
=-ieD^\gamma\,\ ,\\
\label{vz}
\Gamma_{eff}^{Z \sq{t}{}\sq{t}{}}&=&-\frac{ie}{s_Wc_W}\left[\left( \ff\right)
\left(1+\dz{\sq{t}{i}}{}-\dz{2}{Z}\right)\
-\ \sin\theta_{\tilde{t}}\cos\theta_{\tilde{t}}\frac{(\dz{12}{\tilde{t}}+
\dz{21}{\tilde{t}})}{2}\ +\ \Lambda^Z\ \right]\nonumber\\
&=&\ -\frac{ie}{s_Wc_W}D^Z 
\enea
where $\Lambda^\gamma$ and $\Lambda^Z$ are the vertex corrections by
exchanging virtual Higgs bosons, charginos and neutralinos to the
$\gamma$ and $Z$ vertices respectively, as depicted in Fig. 5.
It should be noted that the contribution from Fig. 5(f) is zero.
The contributions coming from Fig. 5(d)
and 5(e) cancel each other because the $\sq{q}{}-\sq{q}{}-A^0$ coupling 
changes signs when the momentum of the squark changes signs. Our 
results are analytically and numerically confirmed by that the UV divergence
are cancelled precisely as it should be. 
The results have been verified by testing the Ward identity.
The analytic expressions for $\Lambda^\gamma$ and $\Lambda^Z$ are 
listed in Appendix B.
\subsection{cross section}
We need to consider two types of contributions, one is the vertex corrections and the other is the corrections to the internal propagators shown in (\ref{pro2}) and (\ref{pro}). We denote the amplitude due to the $\gamma Z$ mixing as $T^{\gamma Z}$ and $T^{Z \gamma}$ where $T^{\gamma Z}$ corresponds to the photon propagator on the left and the $Z$ boson propagator on the right in Fig. 2 and $T^{\gamma Z}$ the vice versa. We use $T^\gamma$ and $T^Z$ to represent the amplitudes calculated from (\ref{vg}) and (\ref {vz}).
Then the cross section can be expressed as
\begin{eqnarray}
\sigma&=&\frac{\pi\alpha^2}{s} \left( 1-\frac{4\mass{t}{i}}{s} \right)^{\frac{3}{2}}\cdot \left[L+M\cdot{ s^2 \over {(s-M_Z^2)}^2}+N\cdot{ s \over {(s-M_Z^2)}}\right]
\end{eqnarray}
%\left( D^\gamma \right)^2\cdot \frac{8}{s}\right.\ \nonumber\\
%&&+\ \
%\left.\left(\frac{D^Z}{s^2c^2}\right)^2\frac{s}{(s-M_Z^2)^2}(1-4s^2+8s^4)
%\ +\ \frac{8}{3}\frac{D^Z D^\gamma}{s^2c^2}\frac{1}{s-M_Z^2}(1-4s^2)
%\right]
where $s=(p_1+p_2)^2$ is the s-channal Mandelstam variable(Fig. 1) and $L$, $M$ and $N$ are
\bea
L&=&8\left( D^\gamma \right)^2\, ,\nonumber\\
M&=&(1-4s_W^2+8s_W^2)\left[{\left({D^Z} \over {s_W^2c_W^2}\right)}^2
+{{\delta Z}^{\gamma Z}
{D^\gamma}{D^Z} \over 4s_W^3c_W^3}\right]+{\delta Z^{\gamma Z}{({D^Z})}^2 \over 2s_W^3c_W^3}(1-4s_W^2)\, ,\nonumber\\
N&=&{4D^ZD^{\gamma} \over s_W^2c_W^2}(1-4s_W^2)
+{\delta Z^{\gamma Z}{\left(D^{\gamma}\right)}^2 \over 2c_Ws_W}(1-4s_W^2)
+{2\delta Z^{\gamma Z}D^{Z}D^{\gamma} \over s_Wc_W}\, .\nonumber
\enea
$L$ comes from the square of $T^\gamma$. The first term in the square bracket of $M$ comes from the square of $T^Z$ while the second term in it comes from the interference between $T^{Z \gamma}$ and $T^{Z}$. The last term in $M$ comes from the interference between $T^{\gamma Z}$ and $T^{Z}$. The first term in $N$ is due to the interference between $T^\gamma$ and $T^Z$, the second term is due to the interfernce between $T^{Z \gamma}$ and $T^{\gamma}$ and the last term is due to the interference between $T^{\gamma Z}$ and $T^{\gamma}$.

Throwing away the one-loop corrections, we 
regain the tree-level formula.
%The above expression dose not contain the $\gamma-Z$ mixing
%terms which must be given separately. They can be easily written out
%by using Eq.~\ref{pro} as internal propagator.
%The expressions for $\delta\sigma$ coming from $\gamma -Z$ mixing
%can be written as
%\be
%\delta\sigma^{\gamma Z}=\frac{\pi}{8}\alpha^2 
%\ene
\subsection{Higgs boson mass formula}
%There are two scalar doublets in the MSSM which implies five physical
%Higgs bosons: two scalars $h^0$, $H^0$, one pseudoscalar $A^0$, and
%two charged ones $H^\pm$\cite{mssm}. 
The Higgs sector is strongly
constrained by supersymmetry\cite{higgs}.
%and can be parameterized
%in terms of $\tan\beta$ and the pseudoscalar mass $M_{A^0}$. At the tree
%level 
In the tree level a light Higgs boson exists with an upper mass bound $M_Z$.
Radiative corrections can considerably change the Higgs mass spectrum.
In our calculations we adopt an approximate Higgs mass formula which
incorporates the one loop radiative corrections. It is given by\cite{hmass}
\bea
M_{H^0,h^0,eff}^2&=&\frac{M_{A^0}^2+M_Z^2+\omega_t}{2}\nonumber\\
&&\pm\sqrt{\frac{\left( M_{A^0}^2+M_Z^2 \right)^2+{\omega_t}^2}{4}
-M_{A^0}^2M_Z^2\cos^22\beta+\frac{\omega_t\cos 2\beta}{2}(M_{A^0}^2-M_Z^2)}
\qquad 
\enea
where
\bea
\omega_t&=&\frac{ N_cG_Fm_t^4 }{ \sqrt{2}\pi^2\sin^2\beta }\left(\log
\frac{ m_{\sq{t}{1}}m_{\sq{t}{2}} }{m_t^2}+\frac{ A_t(A_t+\mu\cot\beta) }{ m_{\sq{t}{1}}^2-
m_{\sq{t}{2}}^2 }\log\frac{ m_{\sq{t}{1}}^2 }{ m_{\sq{t}{2}} }\right.\nonumber\\
&&+\left.\frac{ A_t^2(A_t+\mu\cot\beta)^2 }{ (m_{\sq{t}{1}}^2-m_{\sq{t}{2}}^2)^2 }
\left( 1-\frac{ m_{\sq{t}{1}}^2+m_{\sq{t}{2}}^2 }{ m_{\sq{t}{1}}^2-m_{\sq{t}{2}}^2 }
\log\frac{ m_{\sq{t}{1}} }{ m_{\sq{t}{2}} }\right)\right)\quad .
\enea
Let $\omega_t=0$ we return to the tree level formula of neutral Higgs boson masses.
Although the correction about Higgs masses is a two-loop effect
to the squark pair production, we find it can greatly affect the numerical
results. 
%If we instead use the tree level formula the $M_{h^0}$ is 
%too small in most parameter space and the final corrections to the
%production rates will be too large.

The charged Higgs boson mass is given by 
\be M_{H^\pm}^2=M_{A^0}^2+M_W^2\, .
\ene

\section {NUMERICAL RESULTS}
Now we turn to discuss the numerical results. 
Since the cross section of squark pair production is 
sensitive to the squark masses, we use the two stop masses,
$m_{\sq{t}{1}}$ and $m_{\sq{t}{2}}$,
as input parameters. 
%The stop and sbottom
%mass matrices in interaction eigenstates are given by 
%\bea
%m_{\tilde{t}}^2& =& \left(\begin{array}{cc}
%m_{\tilde{Q}}^2+m_t^2+\cos 2\beta(0.5 - \frac{2}{3}s^2)M_Z^2&
%- m_t(\mu\cot\beta + A_t)\\
%- m_t(\mu\cot\beta + A_t)&
%m_{\tilde{t}}^2+m_t^2+\frac{2}{3}s^2\cos 2\beta 
%\end{array}
%\right)\quad ,\nonumber\\
%m_{\tilde{b}}^2& =& \left(\begin{array}{cc}
%m_{\tilde{Q}}^2+m_b^2-\cos 2\beta(0.5 + \frac{1}{3}s^2)M_Z^2&
%- m_b(\mu\tan\beta + A_b)\\
%- m_b(\mu\tan\beta + A_b)&
%m_{\tilde{b}}^2+m_b^2-\frac{1}{3}s^2\cos 2\beta
%\end{array}
%\right)
%\enea
%Then the combination $\mu\cot\beta + A_t$ can be determined from 
%$m_{\sq{t}{1}}$ and $m_{\sq{t}{2}}$.
Making the following assumptions for simplicity
\bea
m_{\sq{Q}{L}}&=&m_{\sq{t}{R}}\ =\ m_{\sq{b}{R}}\,\ ,\nonumber\\
A_t&=&A_b\,\ ,
\enea
where the $m$s and $A$s are the scalar masses and  trilinear
soft breaking parameters, 
we are left with only two free parameters  
in the squark sector.
The sbottom masses are then determined by $m_{\sq{t}{1}}$, $m_{\sq{t}{2}}$, $\mu$ and $\tan\beta$. For simplicity we also assume the
GUT relation $m_1=(5/3)\tan^2\theta_W m_2$ where $m_1$ and $m_2$ are
U(1) and SU(2) gaugino masses respectively. The chargino and neutralino
sectors are determined by taking $m_2$ as another free parameter. $m_{A^0}$
and $\tan\beta$ determine the MSSM Higgs sector. These free parameters
are constrained by the experimental mass bounds. We impose 
$m_{h^0}>90GeV$\cite{h0mass}, $m_{\chi^0_1}>35GeV$, $m_{\chi^+_1}>95GeV$\cite{char,pdg}
and $m_{\sq{b}{1}}>150GeV$. To discuss the large Yukawa couplings we focus our
attention on the regions of small and large $\tan\beta$. The MSSM may seem unnatural for these values of $\tan\beta$\cite{natu}.
However, 
they are actually not
excluded by present experiments even for $m_{h^0} \ge 90GeV$(See e.g. Ref.~\cite{tan}). 
%Then all the four squark masses $m_{\sq{t}{1,2}},\, m_{\sq{b}{1,2}}$
%are determined by fixing $\tan\beta$ and $\mu$. Adding another
%parameter, the pseudoscalar Higgs boson mass $m_{A^0}$, we get
%the whole parameter space we need in our calculation.
%We find that the electroweak correction ${\delta \sigma}/{\sigma}
%=(\sigma^{cor}-\sigma^{tree})/\sigma^{tree}$
%can be very large for taking large $\mu$. This is reasonable for the
%combination $\mu-A_t\cot\beta$ appearing in the interaction vertex(see Appendix A)
%can be very large for taking large $\mu$ while the combination
%$\mu\cot\beta+A_t$ is fixed. However, large $\mu$ may make one of the two
%sbottoms too light. We always require that $m_{\sq{b}{1}} > 150GeV$ which
%constrains the scope of $\mu$.
Other parameters are taken as
$\alpha=1/128$, $M_W=80.4GeV$, $M_Z=91.2GeV$, $m_t=174GeV$, $m_b=4.7GeV$ and
$\sin^2\theta_W=0.223$.

In Fig. 6, we show the cross section $\sigma (\ee\rightarrow
\sq{t}{i}{\bar{\sq{t}{i}}},\sq{b}{i}{\bar{\tilde{b}}_{i}})$ as a function of
the collision energy $\sqrt{s}$ for $m_{\sq{t}{1}}=150GeV$,
$m_{\sq{t}{2}}=450GeV$ and $\mu=m_{A^0}=m_2=400GeV$ for small and large
$\tan\beta$ scenarios. For $\tan\beta=1.5$ the two sbottom quarks
are almost degenerate. However, for $\tan\beta=30$ the lighter sbottom
can be as light as $\sq{t}{1}$ and its production rates are much larger
than those of the heavier one.

We then calculate the corrections to the cross section 
$\sigma (\ee\rightarrow\sq{t}{1}\sq{t}{1})$ at $\sqrt{s}=206GeV$
at which LEP2 can run in 2000~\cite{lep2000}.
In Fig. 7, we show ${\delta \sigma}/{\sigma}$
as a function of the parameter $\mu$ by taking $m_{\sq{t}{1}}=92GeV$, which
is slightly heavier than the present lower limit\cite{pdg,squ}, and
$m_{\sq{t}{2}}=350GeV$ for $\tan\beta=1.5$,
$m_{A^0}=400,800GeV$ and $m_2=200,800GeV$. We can see that the
corrections are not sensitive to $m_2$.
For large $\mu$
the corrections can be quite large, which is reasonable since $\mu$ directly enters
the Higgs boson and squark coupling vertices. They are generally larger than the SUSY-QCD
corrections due to gluino exchanges\cite{eesq}.
Fig. 8 shows that $\delta\sigma/\sigma$ is also sensitive to $m_{A^0}$
and is more sensitive for smaller $\tan\beta$.

In Figs. 9--11, we present $\delta\sigma/\sigma$ 
for $\sqrt{s}=500GeV$,
$m_{\sq{t}{1}}=150GeV$, $m_{\sq{t}{2}}=450GeV$. For these mass values of the stops 
$\sqrt{s}=500GeV$ is close to the peak for $\sq{t}{1}$ pair production and also to that for
$\sq{b}{1}$ pair production at large $\tan\beta$ as shown in Fig. 6.
Fig. 9 shows $\delta\sigma/\sigma$ 
as a function of $\mu$ for $\tan\beta=1.5$ and Fig. 10 shows
$\delta\sigma/\sigma$ 
as a function of $\mu$ for $\tan\beta=0.6$. We can see that for
$\tan\beta < 1$ the cross section for stop production is greatly
suppressed. We find the cusps in the two figures are a threshold effect mainly coming from Fig. 5(i) when $m_{\chi^-}\approx 250GeV$. 
Fig. 11 shows the correction as a function of $m_{A^0}$ for $\mu=400GeV$
and several values of $\tan\beta$. From this figure we also see that 
the corrections are negative and the cross section
%for $\mu > 400GeV$
%or $m_{A^0}$ 
is suppressed for $\tan\beta < 1$.

We then discuss a scenario with large SUSY parameters at
$\sqrt{s}=2000GeV$. We take $m_{\sq{t}{1}}=400GeV$ and 
$m_{\sq{t}{2}}=800GeV$ in the following discussions.
Figs. 12 and 13 show the ratio of the corrections to the tree level result for $\sigma(\ee\rightarrow
\sq{t}{1}\bar{\sq{t}{1}},\sq{t}{2}\bar{\sq{t}{2}})$ 
as a function of $\mu$ and $m_{A^0}$
respectively when $m_2=1000GeV$. The cusp in Fig. 13   
mainly comes from Fig. 5(c) when $m_{H^+}\approx 1000GeV$. The corrections
are large for small $\tan\beta$. The corrections for $\sigma(\ee\rightarrow
\sq{t}{2}\bar{\sq{t}{2}})$ show a singularity which stems from the wave
function renormalization for $\sq{t}{2}$ at $m_{A^0}=400GeV$, 
where $m_{{\tilde t}_2}=m_{{\tilde t}_1}+m_{A^0}$. Such singularity was also mentioned
in~\cite{eesq,sing}.
The corrections can even reach up to 20\% in this case.

In Fig. 14 we give the corrections to the sbottom production rates
for small and large $\tan\beta$ scenarios. When $\tan\beta=30$
the corrections for $\sq{b}{1}$ and $\sq{b}{2}$ are both positive
and those for $\sq{b}{2}$ can be larger than 20\%. For $\tan\beta=2$ 
the corrections tend to increase the $\sq{b}{2}$ production and
decrease the $\sq{b}{1}$ production.

Finally, we will compare the contributions from three classes of
diagrams, {\it i.e.}, (1) the vertex and squark wave function 
corrections from Higgs bosons, 
(2) those from charginos and neutralinos, and (3) corrections to 
the gauge boson propagator. 
Especially we will show the importance of the third class.
We find this contribution is sensitive 
to the mass difference between the two scalar top quark mass eigenstates.
In Fig. 15 we give three classes of contributions as functions of parameter
$\mu$ for $\sqrt{s}=1500GeV$, $\tan\beta=3$, $m_{A^0}=400GeV$, $m_2=800GeV$,
$m_{\tilde{t_1}}=300GeV$ and $m_{\tilde{t_2}}=500GeV,\ 800GeV$. In fig. 16
we give the same quantities as functions of parameter $m_{\tilde{t_1}}$
by fixing $m_{\tilde{t_2}}=800GeV$, $\mu=-300GeV$ and take other parameters
the same as those in Fig. 15. We find that the contribution from corrections to
gauge boson propagator can be as large as about 7\% of the total cross section.
It can be seen from these figures that the contribution from corrections to
the gauge boson propagator is larger than those from charginos
and neutralinos in a large region of the parameter space and it is opposite in sign
and not much smaller than the contributions from Higgs bosons.

\section {SUMMARY AND CONCLUSION}
In summary, we have calculated the large Yukawa coupling corrections
to the diagonal stop and sbottom pair production 
in $\ee$ annihilation.
We include also terms of the self-energy corrections to gauge bosons enhanced by large masses.
% due to large Yukawa couplings.
 We discuss
the corrections as functions of different SUSY parameters. They
are found to be quite significant and are larger than
the SUSY-QCD corrections by gluino exchanges\cite{eesq}
in a large region of the MSSM parameter space. They can even be
comparable to the conventional QCD corrections which is about 20\%\cite{eesq}.
The corrections can be both
positive and negative. We find the corrections
are quite sensitive to the parameters $\mu$, $m_{A^0}$ and $\tan\beta$.
They are not sensitive to the gaugino mass $m_2$. 
%They are not
%sensitive to $\tan\beta$ for $\tan\beta > 1$  but very
%sensitive to it for $\tan\beta < 1$. 
In conclusion,
when one consider the
%with $m_{A^0} < 400 GeV$ or $\tan\beta < 1$
%or $\mu > 400 GeV$ scenarios 
third generation squark production in the MSSM such corrections should
not be ignored if precision prediction is needed.

After we finished the work we became aware of the work of
H. Eberl {\it et al}\cite{eberl} 
in which a large part of this work had been done. However, they
did not include the contributions coming from corrections to the gauge boson
propagator. As discussed at the end of the last section this contribution
is sizable in a large region of parameter space and should be taken into
account for consistency. The corrections to the gauge boson propagator has
been calculated in a different renormalization scheme in connection to
the process of chargino pair production\cite{diaz}. The effects of
charginos, neutralinos and Higgs bosons in the loop are not considered in
that paper. Apart from this difference we agree with the 
analytical formulas given in their paper. Our numerical results contain studies
of effects of varying different supersymmetric parameters which are not given in
ref\cite{eberl}. We have checked that by taking the same parameter values as in
\cite{eberl} and neglecting the corrections to the gauge boson propagator we
obtain numerical results close to theirs.
%Omitting
%the gauge boson contributions we find our results are in agreement with
%those of H. Eberl's both analytically and numerically.
\section*{ACKNOWLEDGMENTS}
Y-B. Dai's work is supported by the National Science Foundation of China.
\begin{center}
{\bf Appendix A}\\
\end{center}

\setcounter{num}{1}
\setcounter{equation}{0}
\def\theequation{\Alph{num}.\arabic{equation}}
\newcommand{\Aij}{{Z_{\tilde t}^{1i} \over c_W}({Z_N^{1j}s_W \over 3}+Z_N^{2j}c_W)+{m_t \over {M_W \sin \beta}}Z_{\tilde t}^{2i}Z_N^{4j}}
\newcommand{\Bij}{{-4\tan\theta_W \over 3}Z_{\tilde t}^{2i}Z_N^{1j}+{m_t \over {M_W \sin \beta}}Z_{\tilde t}^{1i}Z_N^{4j}}
\newcommand{\Cij}{-\sqrt{2}Z_{\tilde t}^{1i}{Z^+}^{1j}+{m_t \over {M_W \sin \beta}}Z_{\tilde t}^{2i}{Z^+}^{2j}}
\newcommand{\Dij}{{m_b \over {M_W \cos \beta}}Z_{\tilde t}^{2i}{Z^-}^{2j}}
\newcommand{\Eij}{{Z_{\tilde b}^{1i} \over c_W}({Z_N^{1j}s_W \over 3}-Z_N^{2j}c_W)+{m_b \over {M_W \cos \beta}}Z_{\tilde b}^{2i}Z_N^{3j}}
\newcommand{\Fij}{{2\tan\theta_W \over 3}(Z_{\tilde b}^{2i}Z_N^{1j})+{m_b \over {M_W \sin \beta}}Z_{\tilde b}^{1i}Z_N^{3j}}
\newcommand{\Gij}{-\sqrt{2}Z_{\tilde b}^{1i}{Z^-}^{1j}+{m_b \over {M_W \cos \beta}}Z_{\tilde b}^{2i}{Z^-}^{2j}}
\newcommand{\Hij}{{m_t \over {M_W \sin \beta}}Z_{\tilde b}^{1i}{Z^+}^{2j}}

\newcommand{\AAij}[2]{R_{#1#2}}
\newcommand{\BBij}[2]{S_{#1#2}}
\newcommand{\CCij}[2]{U_{#1#2}}
\newcommand{\DDij}[2]{V_{#1#2}}
\newcommand{\EEij}{{m_b \over {M_W \cos \beta}}Z_{\tilde b}^{i}Z_N^{3j}}
\newcommand{\FFij}{{m_b \over {M_W \sin \beta}}Z_{\tilde b}^{i}Z_N^{3j}}
\newcommand{\GGij}{{m_b \over {M_W \cos \beta}}Z_{\tilde b}^{i}{Z^-}^{2j}}
\newcommand{\HHij}{{m_t \over {M_W \sin \beta}}Z_{\tilde b}^{i}{Z^+}^{2j}}
\def\plb  #1 #2 #3 #4 {Phys.~Lett.        {\bf B#1}, #2 (#3)#4 }
\def\epjc #1 #2 #3 #4 {Eur.~Phys.~J.      {\bf C#1}, #2 (#3)#4 }
In this appendix we list the relevant pieces of the SUSY Lagrangian 
in terms of the mass eigenstates. 
We follow the conventions of ref\cite{comp}, where the full Lagrangian
and the complete set of Feynman rules for the MSSM are given. Some abbreviations of the vertex couplings are defined here, which will appear in the analytic expressions in next appendix.

\bea
{\cal L}_{A^0\sq{t}{i}^\ast\sq{t}{j}}&=& (i-j)\cdot\frac{g_2m_t}{2M_W}
(\mu-A_t\cot\beta)\, \, , \\
{\cal L}_{G^0\sq{t}{i}^\ast\sq{t}{j}}&=&(j-i)\cdot\frac{g_2m_t}{2M_W}
(\mu\cot\beta+A_t)\, \, , \\
{\cal L}_{H_k^0{{\sq{t}{i}}^\ast}\sq{t}{j}}&=&-ig_2\left[
\frac{2}{3}M_W\tan^2\theta_WB_R^k\left(
\delta^{ij}+\frac{3-8\ s_W^2}{4s_W^2}Z_{\tilde t}^{1i}Z_{\tilde t}^{1j}\right)+
\frac{m_t^2}{M_W\sin\beta}Z_R^{2k}\delta^{ij}\right.\nonumber	\\
&&\left.-\frac{m_t}{2M_W\ s_W}(Z_{\tilde t}^{1i}Z_{\tilde t}^{2j}+Z_{\tilde t}^{1j}Z_{\tilde t}^{2i})
(A_tZ_R^{2k}+\mu Z_R^{1k})\right]\, \nonumber\\
& = & -ig_2\Gamma _{ijk}\, \, ,\\
{\cal L}_{H^+{{\sq{t}{i}}^\ast}\sq{b}{j}}&=&ig_2\left[\frac{1}{\sqrt{2}}\left(
-M_W\sin 2\beta+\frac{m_b^2}{M_W}\tan\beta+\frac{m_t^2}{M_W}\cot\beta\right)
Z_{\tilde b}^{1j}Z_{\tilde t}^{1i}+\frac{m_tm_b}{\sqrt{2}M_W\ s_W\ c_W}Z_{\tilde b}^{2j}Z_{\tilde t}^{2i}\right.\nonumber\\
&&\left.+(\mu-A_t\cot\beta)\frac{m_t}{\sqrt{2}M_W}Z_{\tilde b}^{1j}Z_{\tilde t}^{2i}+
(\mu-A_b\tan\beta)\frac{m_b}{\sqrt{2}M_W}Z_{\tilde b}^{2j}Z_{\tilde t}^{1i}\right]\, \nonumber\\
&=&ig_2D_{ij}\, \, ,\\
{\cal L}_{G^+{{\sq{t}{i}}^\ast}\sq{b}{j}}&=&ig_2\left[\frac{1}{\sqrt{2}}\left(
M_W\cos 2\beta+\frac{m_t^2}{M_W}-\frac{m_b^2}{M_W}\right)
Z_{\tilde b}^{1j}Z_{\tilde t}^{1i}\right.\nonumber\\
%-\frac{m_t}{\sqrt{2}M_W}(\mu\cot\beta+A_t)Z_{\tilde b}^{1j}Z_{\tilde t}{2i}\right.\nonumber\\
&&\left.-(\mu\cot\beta+A_t)\frac{m_t}{\sqrt{2}M_W}Z_{\tilde b}^{1j}Z_{\tilde t}^{2i}+
(\mu\tan\beta+A_b)\frac{m_b}{\sqrt{2}M_W}Z_{\tilde b}^{2j}Z_{\tilde t}^{1i}\right]\, \nonumber\\
&=& ig_2D'_{ij}\, \, ,\\
{\cal L}_{A^0{\sq{b}{i}^\ast}\sq{b}{j}}&=& (i-j)\cdot\frac{g_2m_b}{2M_W}
(\mu-A_b\tan\beta)\, \, , \\
{\cal L}_{G^0\sq{b}{i}^\ast\sq{b}{j}}&=&(j-i)\cdot\frac{g_2m_b}{2M_W}
(\mu\tan\beta+A_b)\, \, , \\
{\cal L}_{H_k^0{{\sq{b}{i}}^\ast}\sq{b}{j}}&=&ig_2\left[
\frac{M_W}{3}\tan^2\theta_WB_R^k\left(
\delta^{ij}+\frac{3-4\ s_W^2}{2\ s_W^2}Z_{\tilde b}^{1i}Z_{\tilde b}^{1j}\right)-
\frac{m_b^2}{M_W\cos\beta}Z_R^{1k}\delta^{ij}\right.\nonumber	\\
&&\left.+\frac{m_b}{2M_W\cos\beta}(Z_{\tilde b}^{1i}Z_{\tilde b}^{2j}+Z_{\tilde b}^{1j}Z_{\tilde b}^{2i})
(A_bZ_R^{1k}+\mu Z_R^{2k})\right]\, \, ,\\
{\cal L}_{{\overline{\chi}_j^0}\sq{t}{i}t}&=&{-ig_2 \over \sqrt{2}}\left[\left(\Aij\right) P_L\right. \nonumber\\
&&\left.+\left(\Bij\right) P_R \right]\nonumber\\
&=&{-ig_2 \over \sqrt{2}}\left[R_{ij}P_L+S_{ij}P_R \right]\, \, ,\\
{\cal L}_{{\overline {\chi}}_j^-{\sq{t}{i}^\ast}b}&=&{ig_2 \over \sqrt{2}}\left[\left(\Cij\right) P_L\right. \nonumber\\
&&\left.+\left(\Dij\right) P_R \right]\nonumber\\
&=&{ig_2 \over \sqrt{2}}\left[U_{ij}P_L+V_{ij}P_R \right]\, \, ,\\
{\cal L}_{{\overline {\chi}_j^0}\sq{b}{i}b}&=&{-ig_2 \over \sqrt{2}}\left[\left(\Eij\right) P_L\right. \nonumber\\
&&\left.+\left(\Fij\right) P_R \right]\, \, ,\\
{\cal L}_{{\overline \chi}_j^+\sq{b}{i}t}&=&{ig_2 \over \sqrt{2}}\left[\left(\Gij\right) P_L\right. \nonumber\\
&&\left.+\left(\Hij\right) P_R \right]\, \, .
\enea
In the above expressions, 
$Z_R$, $Z_N$ and $Z^+$ and $Z^-$ are the mixing matrices for the two neutral CP-even Higgs bosons, neutralinos and charginos respectively.
\bea
Z_R&=&\left(\begin{array}{ll} 
\cos\alpha&-\sin\alpha\\
\sin\alpha&\cos\alpha\end{array}\right),\\
B_R^k&=&\left\{\begin{array}{ll}
\cos (\alpha +\beta),& k\ =\ 1 ,\\
-\sin (\alpha +\beta),& k\ =\ 2, \end{array} \right.\\
\enea
and
\be
\tan 2\alpha=\tan 2\beta\frac{m_{A^0}^2+M_Z^2}{m_{A^0}^2-M_Z^2}\, .\ene
\begin{center}
{\bf Appendix B}\\
\end{center}
\setcounter{num}{2}
\setcounter{equation}{0}
\def\theequation{\Alph{num}.\arabic{equation}}
In this appendix we give some analytic results in our calculations.
The vertex corrections in Eq.~(\ref{vg}) and (\ref{vz}) are given by
\bea
\Lambda^\gamma&=& -\frac{g_2^2}{(4\pi )^2}\Bigg\{\frac{2}{3}\bigg[
\left(\frac{m_t}{2M_W}\right)^2(\mu-A_t\cot\beta)^2\ (C_0+2C_1)[
\mass{t}{i},s,\mass{t}{i},m_{A^0}^2,\mass{t}{\alpha},\mass{t}{\alpha}]\nonumber\\
&&+\left(\frac{m_t}{2M_W}\right)^2(\mu\cot\beta+A_t)^2\ (C_0+2C_1)[
\mass{t}{i},s,\mass{t}{i},M_Z^2,\mass{t}{\alpha},\mass{t}{\alpha}]\nonumber\\
&&+(\Gamma_{i\alpha k}\Gamma_{i\alpha k})\ (C_0+2\ C_1)[
\mass{t}{i},s,\mass{t}{i},m_{H_k^0}^2,\mass{t}{\alpha},\mass{t}{\alpha}]\bigg]\nonumber\\
&&+(D_{ij})^2\ (C_0+2\ C_1)[
\mass{t}{i},s,\mass{t}{i},\mass{b}{j},m_{H^+}^2,m_{H^+}^2]\nonumber\\
&&+(D'_{ij})^2\ (C_0+2\ C_1)[
\mass{t}{i},s,\mass{t}{i},\mass{b}{j},M_W^2,M_W^2]\nonumber\\
&&-\frac{1}{3}(D_{ij})^2\ (C_0+2\ C_1)[
\mass{t}{i},s,\mass{t}{i},m_{H^+}^2,\mass{b}{j},\mass{b}{j}]\nonumber\\
&&-\frac{1}{3}(D'_{ij})^2\ (C_0+2\ C_1)[
\mass{t}{i},s,\mass{t}{i},M_W^2,\mass{b}{j},\mass{b}{j}]\nonumber\\
&&-{2 \over 3}\left[2(\AAij{i}{j}\BBij{i}{j})m_{\chi_j^0}m_t(C_0+2C_1)[\mass{t}{i},s,\mass{t}{i},m_{\chi_j^0}^2,m_t^2,m_t^2] \right.\nonumber\\
&&+\left(\AAij{i}{j}^2+\BBij{i}{j}^2\right)\cdot\nonumber\\
&&\left. \left((m_t^2+m_{\tilde{t}_i}^2+m_{\chi_j^0}^2)C_1+m_{\chi_j^0}^2C_0\right)[\mass{t}{i},s,\mass{t}{i},m_{\chi_j^0}^2,m_t^2,m_t^2]+{1 \over 2}B_0[s,m_t^2,m_t^2]\right] \nonumber\\
&&-\left[2( \CCij{i}{j}\DDij{i}{j}) m_{\chi_j^+}m_b(C_0+2C_1)[\mass{t}{i},s,\mass{t}{i},m_{b}^2,m_{\chi_j^+}^2,m_{\chi_j^+}^2] \right.\nonumber\\
&&+\left(\CCij{i}{j}^2+\DDij{i}{j}^2 \right)\cdot\nonumber\\
&&\left. \left((m_b^2+m_{\tilde{t}_i}^2+m_{\chi_j^+}^2)C_1+m_{b}^2C_0\right)[\mass{t}{i},s,\mass{t}{i},m_b^2,m_{\chi_j^+}^2,m_{\chi_j^+}^2]+{1 \over 2}B_0[s,m_{\chi_j^+}^2,m_{\chi_j^+}^2]\right] \nonumber\\
&&+{1\over 3}\left[2(\CCij{i}{j}\DDij{i}{j})m_{\chi_j^+}m_b(C_0+2C_1)[\mass{t}{i},s,\mass{t}{i},m_{\chi_j^+}^2,m_b^2,m_b^2] \right.\nonumber\\
&&+\left(\CCij{i}{j}^2+\DDij{i}{j}^2 \right)\cdot\nonumber\\
&&\left. \left((m_b^2+m_{\tilde{t}_i}^2+m_{\chi_j^+}^2)C_1+m_{\chi_j^+}^2C_0\right)[\mass{t}{i},s,\mass{t}{i},m_{\chi_j^+}^2,m_b^2,m_b^2]+{1 \over 2}B_0[s,m_b^2,m_b^2]\right]  \Bigg\}\\
\Lambda^Z&=&-\frac{g_2^2}{(4\pi )^2}\Bigg\{
\left(\frac{m_t}{2M_W}\right)^2(\mu-A_t\cot\beta)^2\ F_{\alpha\alpha}(C_0+2C_1)[
\mass{t}{i},s,\mass{t}{i},m_{A^0}^2,\mass{t}{\alpha},\mass{t}{\alpha}]\nonumber\\
&&+\left(\frac{m_t}{2M_W}\right)^2(\mu\cot\beta+A_t)^2\ F_{\alpha\alpha}(C_0+2C_1)[
\mass{t}{i},s,\mass{t}{i},M_Z^2,\mass{t}{\alpha},\mass{t}{\alpha}]\nonumber\\
&&+(\Gamma_{i\alpha k}\Gamma_{i\beta k})\ F_{\alpha\beta}\left(C_0
+2C_1\right)[\mass{t}{i},s,\mass{t}{i},m_{H_k^0}^2,\mass{t}{\alpha},\mass{t}{\beta}]\nonumber\\
%+C_1[\mass{t}{\alpha}\leftrightarrow\mass{t}{\beta}]\right)\nonumber\\
&&+(D_{ij})^2(0.5-s_W^2)(C_0+2\ C_1)[
\mass{t}{i},s,\mass{t}{i},\mass{b}{j},m_{H^+}^2,m_{H^+}^2]\nonumber\\
&&+(D'_{ij})^2(0.5-s_W^2)(C_0+2\ C_1)[
\mass{t}{i},s,\mass{t}{i},\mass{b}{j},M_W^2,M_W^2]\nonumber\\
&&-(D_{ij})^2G_{ij}(C_0+2\ C_1)[
\mass{t}{i},s,\mass{t}{i},m_{H^+}^2,\mass{b}{j},\mass{b}{j}]\nonumber\\
&&-(D'_{ij})^2G_{ij}(C_0+2\ C_1)[
\mass{t}{i},s,\mass{t}{i},M_W^2,\mass{b}{j},\mass{b}{j}]\nonumber\\
&&-{1 \over 2}\left[\left({(Z_{\tilde t}^{2i})}^2-{(Z_{\tilde t}^{1i})}^2\right)Z_N^{4j}Z_N^{4k}(Z_N^{4j}Z_N^{4k}-Z_N^{3j}Z_N^{3k}){\left({m_t \over {m_W \sin\beta}}\right)}^2\left((m_{\chi^0_j}m_{\chi^0_k}-m_t^2-\mass{t}{i})\right.\right.\cdot \nonumber\\
&&\left. \left.C_1 -m_t^2C_0\right)[\mass{t}{i},s,\mass{t}{i},m_t^2,m_{\chi_k^0}^2,m_{\chi_j^0}^2]-{1 \over 2}B_0[s,m_{\chi_k^0}^2,m_{\chi_j^0}]\right]\nonumber\\
&&-{1 \over 2}\left[(\AAij{i}{j}\BBij{i}{j})m_{\chi_j^0}m_t(1-{8 \over 3}s_W^2)(C_0+2C_1)[\mass{t}{i},s,\mass{t}{i},m_{\chi_j^0}^2,m_t^2,m_t^2]\right.\nonumber\\
&&+\left(\AAij{i}{j}^2-{4 \over 3}s_W^2\left(\AAij{i}{j}^2+\BBij{i}{j}^2\right) \right)\cdot\nonumber\\
&&\left(\left((m_{\tilde{t}_i}^2+m_{\chi_j^0}^2)C_1+m_{\chi_j^0}^2C_0\right)[\mass{t}{i},s,\mass{t}{i},m_{\chi_j^0}^2,m_t^2,m_t^2]+{1 \over 2}B_0[s,m_t^2,m_t^2]\right)\nonumber \\
&&\left.+\left({\BBij{i}{j}}^2-{4 \over 3}s_W^2\left({\AAij{i}{j}}^2+{\BBij{i}{j}}^2\right)\right)m_t^2C_1[\mass{t}{i},s,\mass{t}{i},m_{\chi_j^0}^2,m_t^2,m_t^2]\right] \nonumber\\
&&-{1 \over 2}\cos(2\theta_W)\left[2(\CCij{i}{j}\DDij{i}{j})m_{\chi_j^+}m_b(C_0+2C_1)[\mass{t}{i},s,\mass{t}{i},m_{b}^2,m_{\chi_j^+}^2,m_{\chi_j^+}^2]\right. \nonumber\\
&&+\left(\CCij{i}{j}^2+\DDij{i}{j}^2 \right)\cdot\nonumber\\
&&\left. \left((m_b^2+m_{\tilde{t}_i}^2+m_{\chi_j^+}^2)C_1+m_{b}^2C_0\right)[\mass{t}{i},s,\mass{t}{i},m_b^2,m_{\chi_j^+}^2,m_{\chi_j^+}^2]+{1 \over 2}B_0[s,m_{\chi_j^+}^2,m_{\chi_j^+}^2]\right] \nonumber\\
&&+{1 \over 2}\left[(\CCij{i}{j}\DDij{i}{j})m_{\chi_j^+}m_b(1-{4 \over 3}s_W^2)(C_0+2C_1)[\mass{t}{i},s,\mass{t}{i},m_{\chi_j^+}^2,m_b^2,m_b^2]\right.\nonumber\\
&&+\left(\CCij{i}{j}^2-{2 \over 3}s_W^2\left(\CCij{i}{j}^2+\DDij{i}{j}^2\right) \right)\cdot\nonumber\\
&&\left(\left((m_{\tilde{t}_i}^2+m_{\chi_j^+}^2)C_1+m_{\chi_j^+}^2C_0\right)[\mass{t}{i},s,\mass{t}{i},m_{\chi_j^+}^2,m_b^2,m_b^2]+{1 \over 2}B_0[s,m_b^2,m_b^2]\right)\nonumber \\
&&+\left.\left({\DDij{i}{j}}^2-{2 \over 3}s_W^2\left({\CCij{i}{j}}^2+{\DDij{i}{j}}^2\right)\right)m_b^2C_1[\mass{t}{i},s,\mass{t}{i},m_{\chi_j^+}^2,m_b^2,m_b^2]\right] \Bigg\}\, \, .
\enea
The analytic expressions for stop self energies are
\bea
\Sigma^{ij}(p^2)&=&{g_2^2 \over {(4\pi)}^2}\left\{(1-\delta^{ij}){\left(\frac{m_t}{2M_W}\right)}^2{\left(\mu-A_t\cot\beta\right)}^2B_0[p^2,\mass{t}{j},m_{A^0}^2]\right.\nonumber\\
&&+(1-\delta^{ij}){\left(\frac{m_t}{2M_W}\right)}^2{\left(\mu\cot\beta+A_t\right)}^2B_0[p^2,\mass{t}{j},m_{Z}^2]\nonumber \\
&&+(\Gamma_{i\alpha k}\Gamma_{j \alpha k})B_0[p^2,\mass{t}{\alpha},m_{H_k^0}^2]+(D_{j\alpha}D_{\alpha i})B_0[p^2,\mass{b}{\alpha},m_{H^+}^2]+(D'_{j\alpha}D'_{\alpha i})B_0[p^2,\mass{b}{\alpha},m_{W}^2]\nonumber\\
&&-2\sin{\theta_{\tilde{t}}}\cos{\theta_{\tilde{t}}}\delta^{ij}\left({3+2s^2_W \over 12 c^2_W}\cos({2\beta})-{m_t^2\cot\beta \over 2m_W^2}+{m_b^2\tan^2 \beta \over 2m_W^2}\right)A_0[m_{H^{\pm}}^2]\nonumber\\
&&+\sin{\theta_{\tilde{t}}}\cos{\theta_{\tilde{t}}}{3-8s_W^2 \over 12 s_W^2}\delta^{ij}\left(\cos(2\beta)A_0[m_{A^0}^2]+\cos(2\alpha)(A_0[m_{H^0}^2]-A_0[m_{h^0}^2])\right)\nonumber\\
&&-\left[m_{\chi_k^0}m_t\left(\AAij{j}{k}\BBij{i}{k}+\BBij{j}{k}\AAij{i}{k}\right)B_0[p^2,m_t^2,m_{\chi_k^0}^2]\right.\nonumber\\
&&+\left(\AAij{j}{k}\AAij{i}{k}+\BBij{j}{k}\BBij{i}{k}\right)\left(A_0[m_{\chi_k^0}^2]+m_t^2B_0[p^2,m_t^2,m_{\chi_k^0}^2]\right)\nonumber\\
&&\left.+\left(\AAij{j}{k}\AAij{i}{k}+\BBij{j}{k}\BBij{i}{k}\right)p^2B_1[p^2,m_t^2,m_{\chi_k^0}^2]\right]\nonumber\\
&&-\left[m_{\chi_k^0}m_b\left(\CCij{j}{k}\DDij{i}{k}+\DDij{j}{k}\CCij{i}{k}\right)B_0[p^2,m_b^2,m_{\chi_k^-}^2]\right. \nonumber\\
&&+\left(\CCij{j}{k}\CCij{i}{k}+\DDij{j}{k}\DDij{i}{k}\right)\left(A_0[m_{\chi_k^-}^2]+m_b^2B_0[p^2,m_b^2,m_{\chi_k^-}^2]\right)\nonumber\\
&&\left.\left.+\left(\CCij{j}{k}\CCij{i}{k}+\DDij{j}{k}\DDij{i}{k}\right)p^2B_1[p^2,m_b^2,m_{\chi_k^-}^2]\right]\right\}
\enea
The analytic expressions for the gauge boson mass corrections are
\bea
\dmass&=&\frac{N_c g_2^2 m_t^2}{64\pi^2 M_W^2}-{g_2^2 \over{16\pi^2M_W^2}}\left\{ {\sin^2(\alpha-\beta)}B_{00}[M_Z^2,m_{A^0}^2,m_{H^0}^2] \right. \nonumber \\
&&\left.+{\cos^2(\alpha-\beta)}B_{00}[M_Z^2,m_{A^0}^2,m_{h^0}^2] \right.\nonumber \\
&&-\left[ {\sin^2(\alpha-\beta)}B_{00}[M_W^2,m_{H^+}^2,m_{H^0}^2]+{\cos^2(\alpha-\beta)}B_{00}[M_W^2,m_{H^+}^2,m_{h^0}^2] \right]\nonumber\\
&&-B_{00}[M_W^2,m_{H^+}^2,m_{A^0}^2]+{1\over 2}A_0[{m_{H^+}^2}]\nonumber\\
&&+4{\left[ { {Z_{\tilde t}^{1i}Z_{\tilde t}^{1j}} \over 2}-{{2 s_W^2 \delta^{ij}} \over 3} \right]}^2B_{00}[M_Z^2,m_{\tilde {t}_i}^2,m_{\tilde {t}_j}^2]\nonumber\\
&&-2\left[ {{4 s_W^4} \over 9}+{{3-8s_W^2}\over 12} {(Z_{\tilde t}^{1i})}^2 \right]A_0[m_{\tilde {t}_i}^2]\nonumber\\
&&+4{\left[ { {Z_{\tilde b}^{1i}Z_{\tilde b}^{1j}} \over 2}-{{s_W^2 \delta^{ij}} \over 3} \right]}^2B_{00}[M_Z^2,{m_{\tilde {b}_i}^2,m_{\tilde {b}_j}^2]}\nonumber\\
&&-2\left[{{s_W}^4\over 9}+{{3-4s_W^2}\over 12} {(Z_{\tilde b}^{1i})}^2 \right]A_0[m_{\tilde {b_i}}^2]\nonumber\\
&&-2{(Z_{\tilde b}^{1i}Z_{\tilde t}^{1j})}^2B_{00}[M_Z^2,m_{\tilde {t}_j}^2,m_{\tilde {t}_i}^2]\nonumber\\
&&+{1 \over 2}{(Z_{\tilde t}^{1i})}^2A_0[m_{\tilde{t}_i}^2]+{1 \over 2}{(Z_{\tilde b}^{1i})}^2A_0[m_{\tilde{b}_i}^2]\nonumber\\
&&-{(Z_N^{4i}Z_N^{4j}-Z_N^{3i}Z_N^{3j})}^2\left(-A0[m_{\chi^0_i}^2]-(m_{\chi_i^0}m_{\chi_j^0}+m_{\chi_j^0}^2)B_0[M_Z^2,m_{\chi_i^0}^2,m_{\chi_j^0}^2]\right.\nonumber\\
&&\left.+2B_{00}[M_Z^2,m_{\chi_i^0}^2,m_{\chi_j^0}^2]\right)\nonumber\\
&&+\left({(Z_N^{4i}{Z^{+}}^{2j})}^2+{(Z_N^{3i}{Z^{-}}^{2j})}^2\right)\left(2B_{00}[M_W^2,m_{\chi_j^-}^2,m_{\chi_i^0}^2]-A0[m_{\chi^0_i}^2] \right.\nonumber\\
&&\left.-m_{\chi_j^-}^2B_0[M_W^2,m_{\chi_j^-}^2,m_{\chi_i^0}^2]\right)\nonumber\\
&&\left.-2m_{\chi_i^0}m_{\chi_j^-}(Z_N^{4i}{Z^{+}}^{2j}Z_N^{3i}{Z^{-}}^{2j})B_0[M_W^2,m_{\chi_j^-}^2,m_{\chi_i^0}^2]\right\}
\enea
%The $p^2$ represents the momentum squared of the external line as shown in Fig 4
$\Gamma_{ijk}$, $D_{ij}$, $D'_{ij}$, $R_{ij}$, $S_{ij}$, $U_{ij}$ and $V_{ij}$ in the above expressions are the vertex couplings
defined in Appendix A. In the concrete calculations we only keep the higgsino sector in the $R_{ij}$, $S_{ij}$, $U_{ij}$ and $V_{ij}$. The other two
coupling constants  
\bea
F_{ij}&=&\frac{1}{2}Z_{\tilde t}^{1j}Z_{\tilde t}^{1j}-\frac{2}{3}s^2_W\delta^{ij}\,\ , \\
G_{ij}&=&\frac{1}{2}Z_{\tilde b}^{1j}Z_{\tilde b}^{1j}-\frac{1}{3}s^2_W\delta^{ij}
\enea
are the couplings of $Z$ boson to top squark and bottom squark respectively.
The relevant scalar functions are defined as follows
\begin {eqnarray}
&&A_0(m^2)={(i \pi^2) }^{-1}{(2 \pi \mu)}^{4-D} \int d^Dq {(q^2-m^2)}^{-1},\\
&&B_0(p_1^2,m_0^2,m_1^2)={(i \pi^2) }^{-1}{(2 \pi \mu)}^{4-D} \int d^Dq {[(q^2-
m_0^2)((q+p_1)^2-m_1^2)]}^{-1},\\
&&C_0(p_1^2,p_{12},p_2^2,m_0^2,m_1^2,m_2^2)\nonumber \\
&&\mbox{\hphantom{123}}={(i \pi^2) }^{-1}{(2 \pi \mu)}^{4-D} \int d^Dq {[(q^2-m
_0^2)((q+p_1)^2-m_1^2)((q+p_2)^2-m_2^2)]}^{-1}\,\ ,
%\nonumber \\
%&&D_0(p_1^2,p_{12},p_{23},p_3^2,p_2^2,p_{13},m_0^2,m_1^2,m_2^2,m_3^2)\nonumber
%\\
%&&\mbox{\hphantom{123}}={{(i \pi^2) }^{-1}{(2 \pi \mu)}^{4-D}}\nonumber \\
%&&\mbox{\hphantom{123}}\int d^Dq {[(q^2-m_0^2)((q+p_1)^2-m_1^2)((q+p_2)^2-m_2^2
%)((q+p_3)^2-m_3^2)]}^{-1}
%,\\
%&&DB_0(p_1^2,m_0^2,m_1^2)={\partial B_0(p_1^2,m_0^2,m_1^2) \over \partial {p_1^
%2}},
\end {eqnarray}
in which $p_{ij}={(p_i-p_j)}^2$.

The definitions of the tensor integrals and the relevant decompositions are given
 below
\begin{eqnarray}
T_{\mu_1 \cdot \cdot \cdot \mu_p}(p_1,\cdot \cdot \cdot,p_{N-1},m_0,\cdot \cdot \cdot ,m_{N-1})={{(2 \pi \mu)}^{4-D} \over {i \pi^2}} \int d^Dq {q_{\mu_1} \cdot \cdot \cdot q_{\mu_n} \over D_0 D_1 \cdot \cdot \cdot D_{N-1}},
\end {eqnarray}
with the denominator factors $D_0=q^2-m_0^2,D_i={(q+p_i)}^2-m_i^2$  (i=1,$\cdot
 \cdot \cdot$,N-1) and $T=B,C,D\cdot \cdot \cdot$ corresponding to $N=2,3,4 \cdot \cdot \cdot$.
\begin{eqnarray}
&&B_{\mu}={p_1}_{\mu}B_1,\\
&&B_{\mu \nu}=g_{\mu\nu}B_{00}+{p_1}_{\mu}{p_1}_{\nu}B_{11},\\
&&C_{\mu}={p_1}_{\mu}C_1+{p_2}_{\mu}C_2=\sum\limits_{i=1}^2 p_{i\mu}C_i,\\
&&C_{\mu \nu}=g_{\mu\nu}C_{00}+{p_1}_{\mu}{p_1}_{\nu}C_{11}+{p_2}_{\mu}{p_2}_{\nu}C_{22}+({p_1}_{\mu}{p_2}_{\nu}+{p_2}_{\mu}{p_1}_{\nu})C_{12}\nonumber\\
&&\mbox{\hphantom{123}}=g_{\mu\nu}C_{00}+\sum\limits_{i,j=1}^2{p_i}_{\mu}{p_j}_
{\nu}C_{ij}\, \ .
%,\\
%&&D_{\mu}=\sum\limits_{i=1}^3 p_{i\mu}D_i,\\
%&&D_{\mu\nu}=g_{\mu\nu}D_{00}+\sum\limits_{i,j=1}^3 {p_i}_{\mu}{p_j}_{\nu}D_{ij
%}.
\end {eqnarray}
%The definitions of the mixing matrices of squarks($Z_{\tilde t}$ and $Z_{\tilde b}$) and 
%the mixinig angles($\theta_{\tilde{t}_i}$ and $\theta_{\tilde{b}_i}$) 
%can be found in Sec. II. 

The analytic expressions of $A_0(m^2)$, $B_0(p^2,m_0^2,m_1^2)$ and $B_1(p^2,m_0^2,m_1^2)$ can be easily obtained and the corresponding divergences are:
\begin{eqnarray}
A_0(m^2)&=& m^2\left({2\over 4-D}-\gamma_E+\ln{4\pi}\right)+\cdot\cdot\cdot,\\
B_0(p^2,m_0^2,m_1^2)&=& {2\over 4-D}-\gamma_E+\ln{4\pi}+\cdot\cdot\cdot,\\
B_1(p^2,m_0^2,m_1^2)&=& {-1\over 2}\left({2\over 4-D}-\gamma_E+\ln{4\pi}\right)+\cdot\cdot\cdot.
\end{eqnarray}
$B_{00}$ can be expressed by $A_0$, $B_1$ and $B_0$ as follows
\begin{eqnarray}
B_{00}(p^2,m_0^2,m_1^2)&=&{1\over 6}\left\{A_0(m_1^2)+2m_0^2B_0(p^2,m_0^2,m_1^2)+(p^2+m_0^2-m_1^2)B_1(p^2,m_0^2,m_1^2)\right.\nonumber\\
&&\left.+m_0^2+m_1^2-{p^2\over 3}\right\}
\end{eqnarray}
and we can extract the divergent part
\begin{eqnarray}
B_{00}(p^2,m_0^2,m_1^2)= \left[{1\over 4}(m_1^2+m_0^2)-{1\over 12}p^2\right]\left({2\over 4-D}-\gamma_E+\ln{4\pi}\right)+\cdot\cdot\cdot.
\end{eqnarray}

The contribution of the top-quark loop to the self-energy of the $Z$ gauge boson is 
\begin{eqnarray}
\Sigma_T^Z (k^2)&=&{\alpha \over 4\pi}\left[{4\over 3}({1\over 8}+{4\over 9}s_W^4-{2\over 3}s_W^2)\left(k^2\Delta_t +(k^2+2m_t^2) F(k^2,m_t,m_t)-{k^2 \over 3}\right)\right.\nonumber\\
&&\left.-{3\over 8s_W^2 c_W^2}m_t^2\left(\Delta_t+F(k^2,m_t,m_t)\right)\right]
\end {eqnarray}
where 
\begin{eqnarray}
F(k^2,m_t^2,m_t^2)&=&-\int_0^1 dx\ \ln{x^2k^2-x k^2+m_t^2-i\epsilon \over m_t^2}.
\end{eqnarray}
For $k^2< m_t^2$ $F(k^2,m_t^2,m_t^2)$ can be expanded as a convergent power series in $k^2\over m_t^2$ with $k^2\over 6m_t^2$ as the first term. Therefore, $m_t^2 F(k^2,m_t^2,m_t^2)$ is not large in this region. For $k^2> m_t^2$ the term $\Sigma_t^Z(k^2)-\Sigma_t^Z(M_Z^2)\over k^2-M_Z^2$ is not enhanced either. Therefore $\Sigma_t^Z(k^2)-\Sigma_t^Z(M_Z^2)\over k^2-M_Z^2$ is not large for all values of $k^2$.

\pagebreak

\newpage
\section*{Figure Captions}
\newcounter{FIG}
\begin{list}{{\bf FIG. \arabic{FIG}}}{\usecounter{FIG}}
\item
The tree-level Feynman diagram for the process $\ee\rightarrow \sq{t}{i}
\bar{\sq{t}{i}}$.
\item
Self-energies and counter terms for internal gauge bosons. Note that
each graph represents four different combinations.
\item
Feynman diagrams for gauge boson mass corrections.
\item
Feynman diagrams for self energies of squarks and their mixing at one loop order.
\item
Corrections to vertex $\gamma (Z)\sq{t}{i}\bar{\sq{t}{i}}$
due to Higgs boson, neutralino and chargino exchanges.
\item
The cross section $\sigma(\ee\rightarrow\sq{t}{i}{\bar{\sq{t}{i}}}$,$\sq{b}{i}{\bar{\tilde{b}}}_{i})$
as a function of the collision energy $\sqrt{s}$ for 
$m_{\sq{t}{1}}=150GeV$,
$m_{\sq{t}{2}}=450GeV$, $\mu=m_{A^0}=m_2=400GeV$ and (a)
$\tan\beta=1.5$, (b) $\tan\beta=30$.
\item
Corrections $\delta\sigma/\sigma$ as a function of $\mu$
for $\ee\rightarrow\sq{t}{1}\bar{\sq{t}{1}}$ at $\sqrt{s}=206GeV$ for 
$m_{\sq{t}{1}}=92GeV$, $m_{\sq{t}{2}}=350GeV$, $\tan\beta=1.5$ and
several values of $m_{A^0}$ and $m_2$.
\item
Corrections $\delta\sigma/\sigma$ as a function of $m_{A^0}$
for $\ee\rightarrow\sq{t}{1}\bar{\sq{t}{1}}$ at $\sqrt{s}=206GeV$ for 
$m_{\sq{t}{1}}=92GeV$, $m_{\sq{t}{2}}=350GeV$, $\mu=600GeV$, $m_2=600GeV$ and
different $\tan\beta$.
\item
Corrections $\delta\sigma/\sigma$ as a function of $\mu$
for $\ee\rightarrow\sq{t}{1}{\bar{\sq{t}{1}}}$ at $\sqrt{s}=500GeV$ for 
$m_{\sq{t}{1}}=150GeV$, $m_{\sq{t}{2}}=450GeV$, $\tan\beta=1.5$, $m_2=600GeV$ and
several values of $m_{A^0}$.
\item
Corrections $\delta\sigma/\sigma$ as a function of $\mu$
for $\ee\rightarrow\sq{t}{1}{\bar{\sq{t}{1}}}$ at $\sqrt{s}=500GeV$ for 
$m_{\sq{t}{1}}=150GeV$, $m_{\sq{t}{2}}=450GeV$, $\tan\beta=0.6$, $m_2=600GeV$ and
several values of $m_{A^0}$.
\item
Corrections $\delta\sigma/\sigma$ as a function of $m_{A^0}$
for $\ee\rightarrow\sq{t}{1}{\bar{\sq{t}{1}}}$ at $\sqrt{s}=500GeV$ for 
$m_{\sq{t}{1}}=150GeV$, $m_{\sq{t}{2}}=450GeV$, $\mu=400GeV$, $m_2=600GeV$ and
several values of $\tan\beta$.
\item
Corrections $\delta\sigma/\sigma$ as a function of $\mu$
for $\ee\rightarrow\sq{t}{1}\bar{\sq{t}{1}},\sq{t}{2}\bar{\sq{t}{2}}$ at $\sqrt{s}=2000GeV$ for 
$m_{\sq{t}{1}}=400GeV$, $m_{\sq{t}{2}}=800GeV$, $\tan\beta=2$, $m_2=1000GeV$ 
and $m_{A^0}=500GeV,900GeV,1300GeV$.
\item
Corrections $\delta\sigma/\sigma$ as a function of $m_{A^0}$
for $\ee\rightarrow\sq{t}{1}\bar{\sq{t}{1}},\sq{t}{2}\bar{\sq{t}{2}}$ at $\sqrt{s}=2000GeV$ for 
$m_{\sq{t}{1}}=400GeV$, $m_{\sq{t}{2}}=800GeV$, $\mu=1200GeV$, $m_2=1000GeV$
and $\tan\beta=1.5,2,30$.
\item
Corrections $\delta\sigma/\sigma$ as a function of $\mu$
for $\ee\rightarrow\sq{b}{1}\bar{\tilde{b}}_{1},\sq{b}{2}\bar{\tilde{b}}_{2}$
at $\sqrt{s}=2000GeV$ for 
$m_{\sq{t}{1}}=400GeV$, $m_{\sq{t}{2}}=800GeV$, $m_2=1000GeV$,
 $m_{A^0}=500GeV,900GeV,1300GeV$ and (a) $\tan\beta=2$, (b) $\tan\beta=30$.
\item
Corrections $\delta\sigma/\sigma$ as a function of $\mu$
for $\ee\rightarrow\sq{t}{1}\bar{\tilde{t}}_{1}$ due to the three
classes of contributions, that is coming from, (1) corrections to the vertex
and squark lines by Higgs bosons (2) by charginos and neutralinos and (3)
corrections to the gauge boson propagators respectively. 
The parameters are taken to be $\sqrt{s}=1500GeV$,
$\tan\beta=3$, $m_{A^0}=400GeV$, $m_2=800GeV$, $m_{\tilde{t_1}}=300GeV$
and $m_{\tilde{t_2}}=500GeV,\ 800GeV$.
\item
Corrections $\delta\sigma/\sigma$ as a function of $m_{\tilde{t_1}}$
for $\ee\rightarrow\sq{t}{1}\bar{\tilde{t}}_{1}$ due to the three
classes of contributions, that is coming from, (1) corrections to the vertex
and squark lines by Higgs bosons (2) by charginos and neutralinos and (3)
corrections to the gauge boson propagators respectively.
The parameters are taken to be $\sqrt{s}=1500GeV$,
$\tan\beta=3$, $m_{A^0}=400GeV$, $m_2=800GeV$, $\mu=-300GeV$
and $m_{\tilde{t_2}}=800GeV$. 
%The Higgs boson and gauge boson contributions
%are sensitive to 
\end{list}
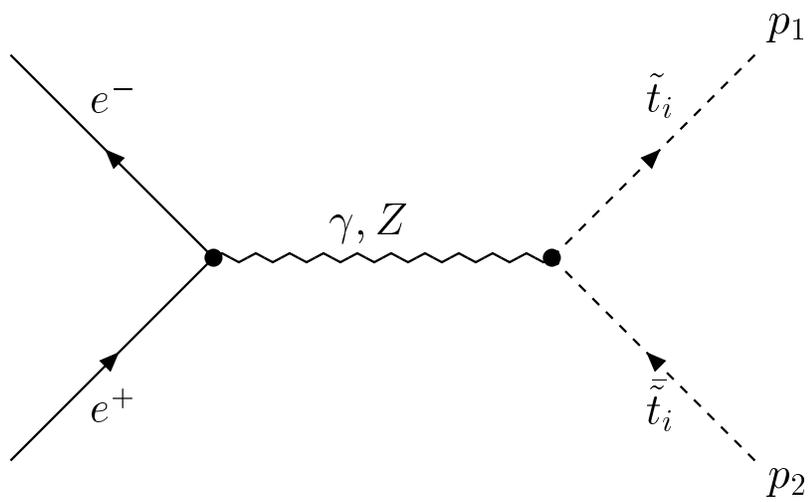
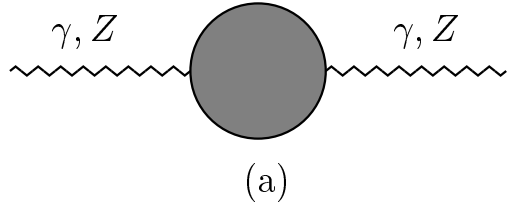
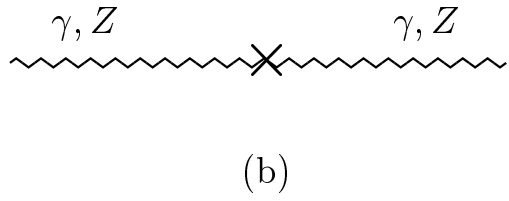
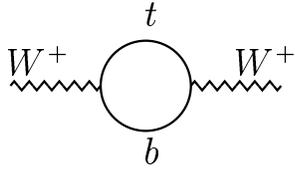
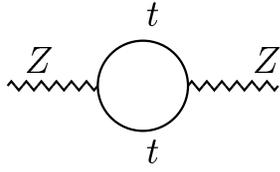
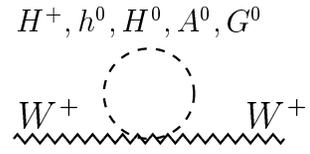
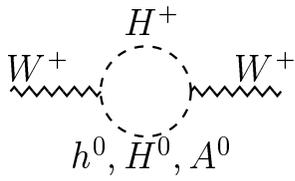
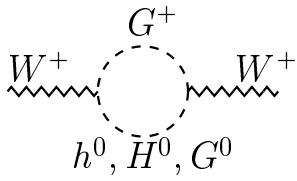
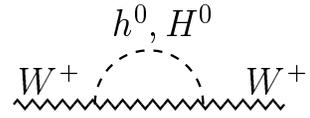
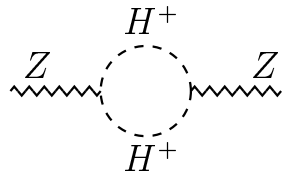
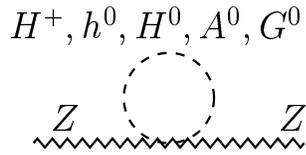
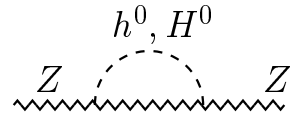
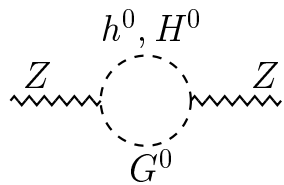
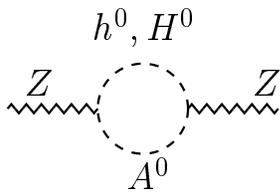
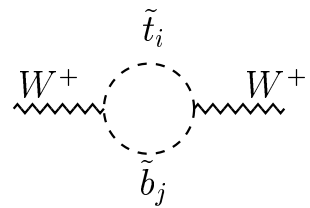
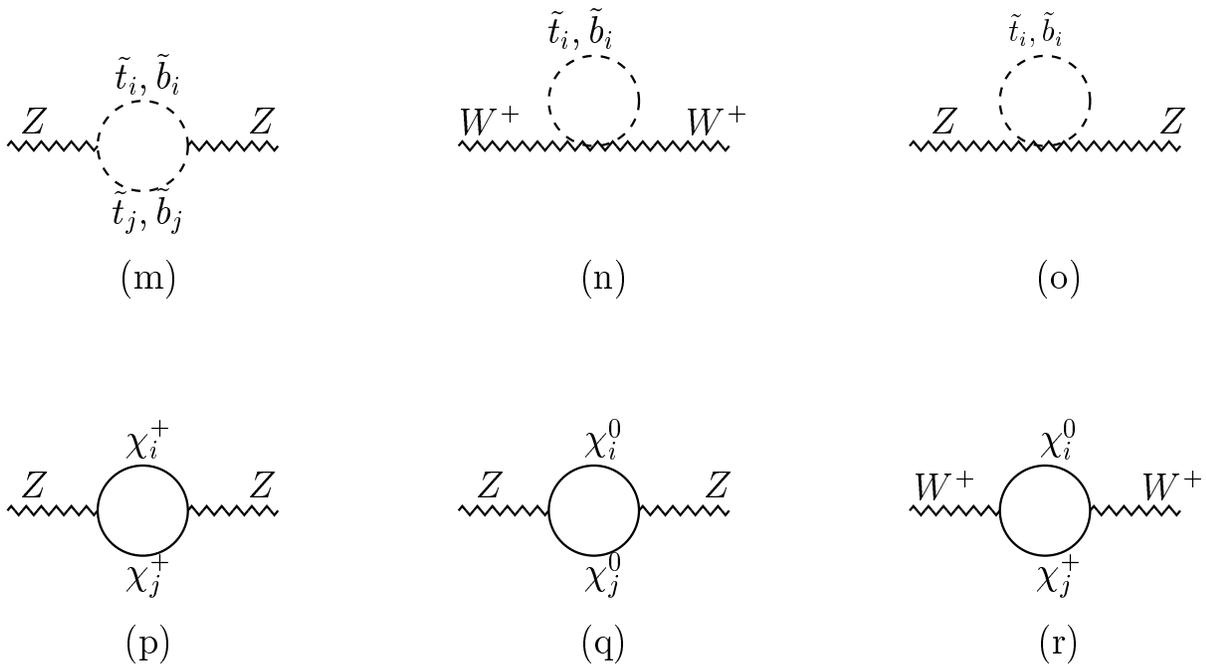
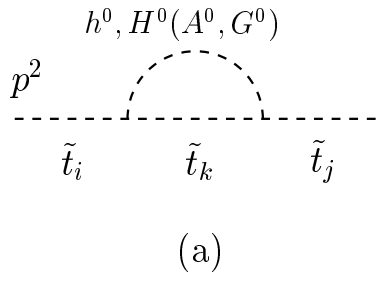
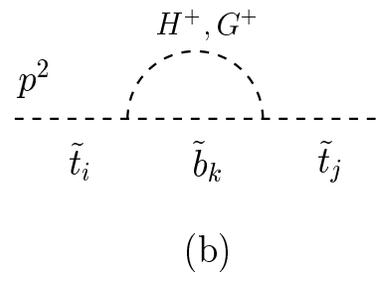
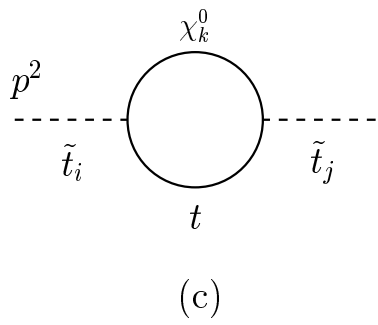
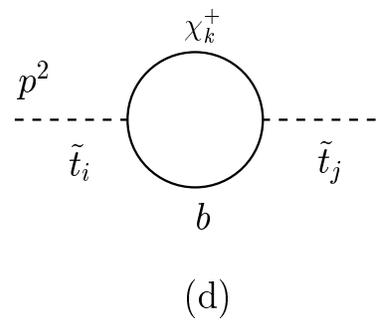
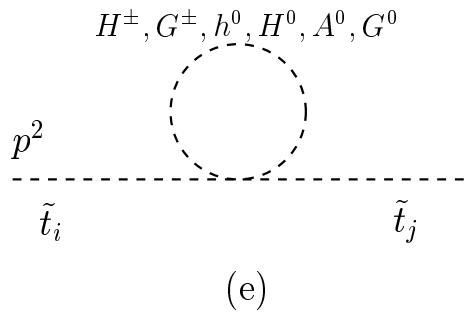
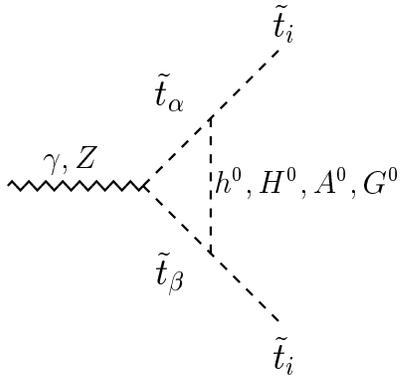
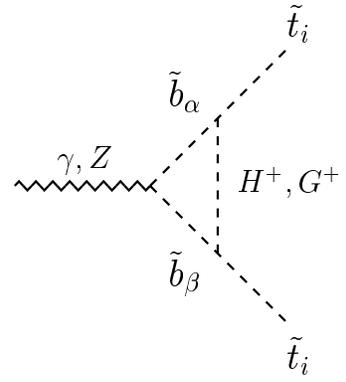
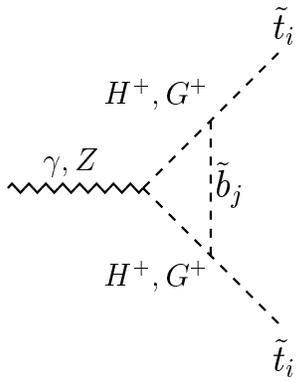
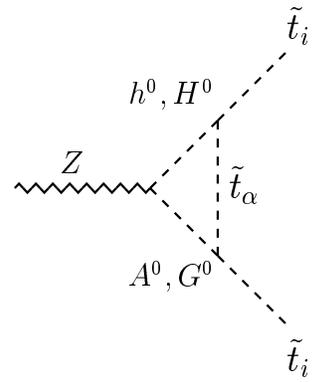
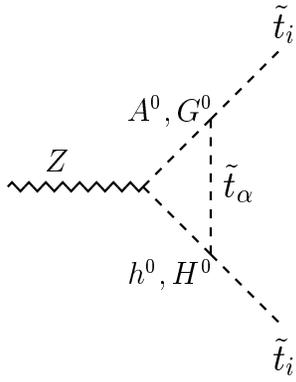
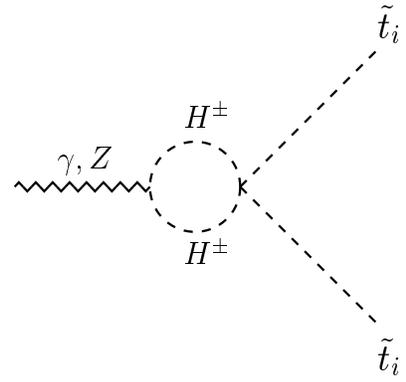
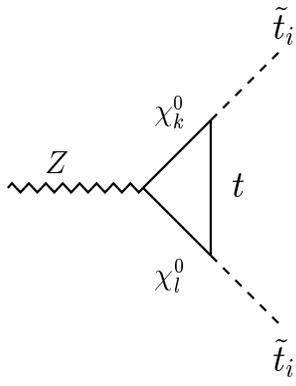
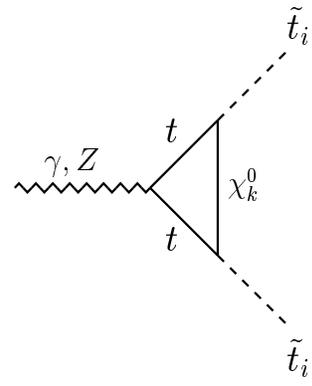
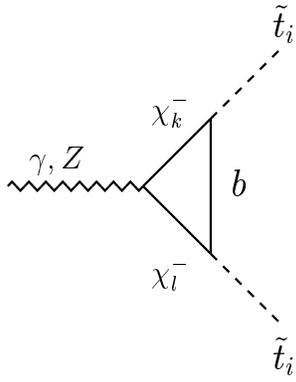
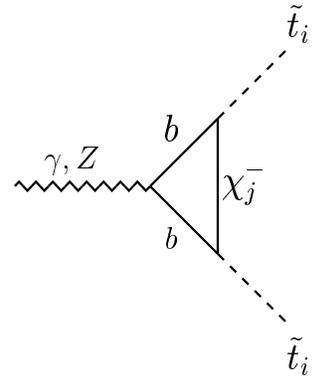
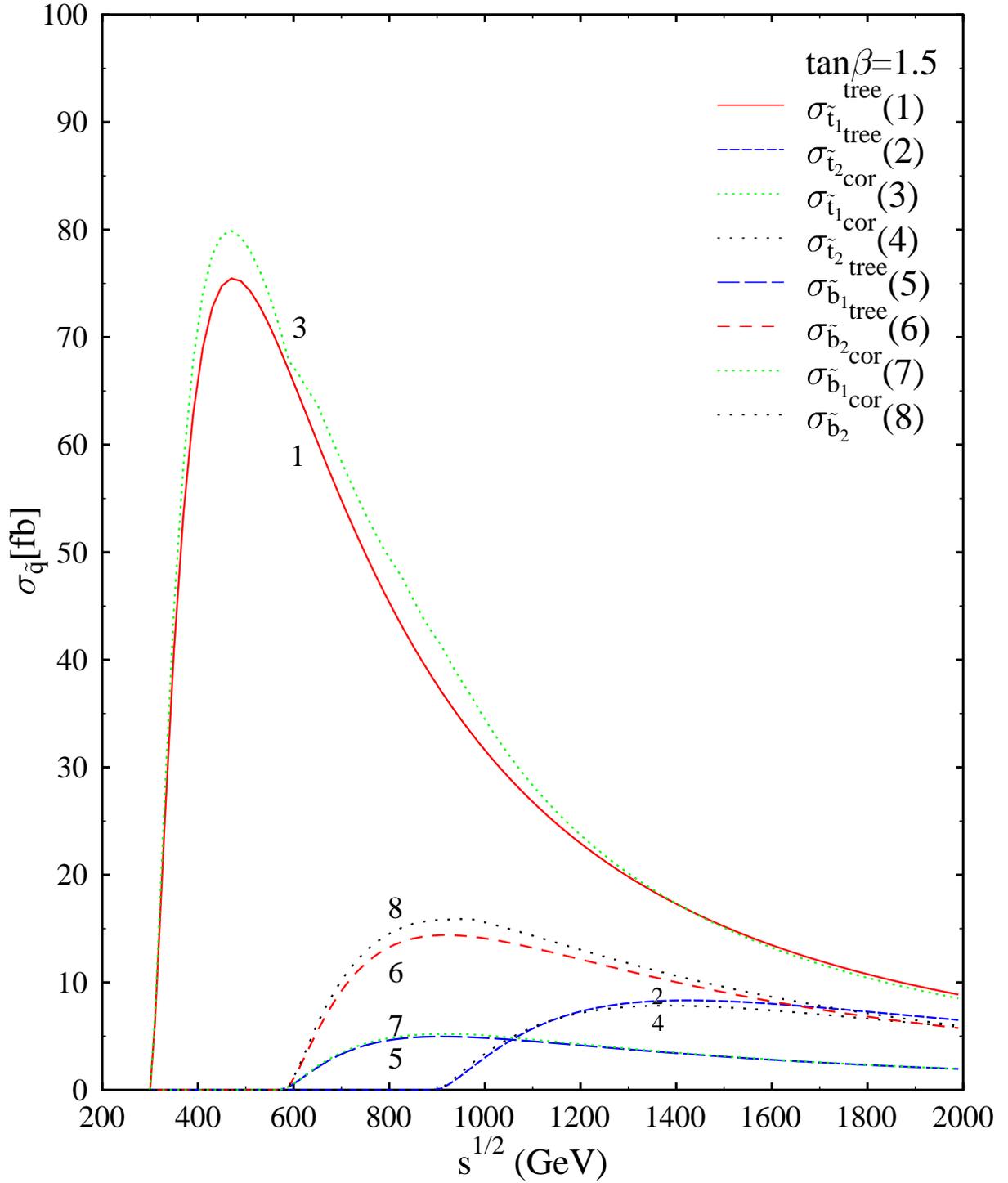
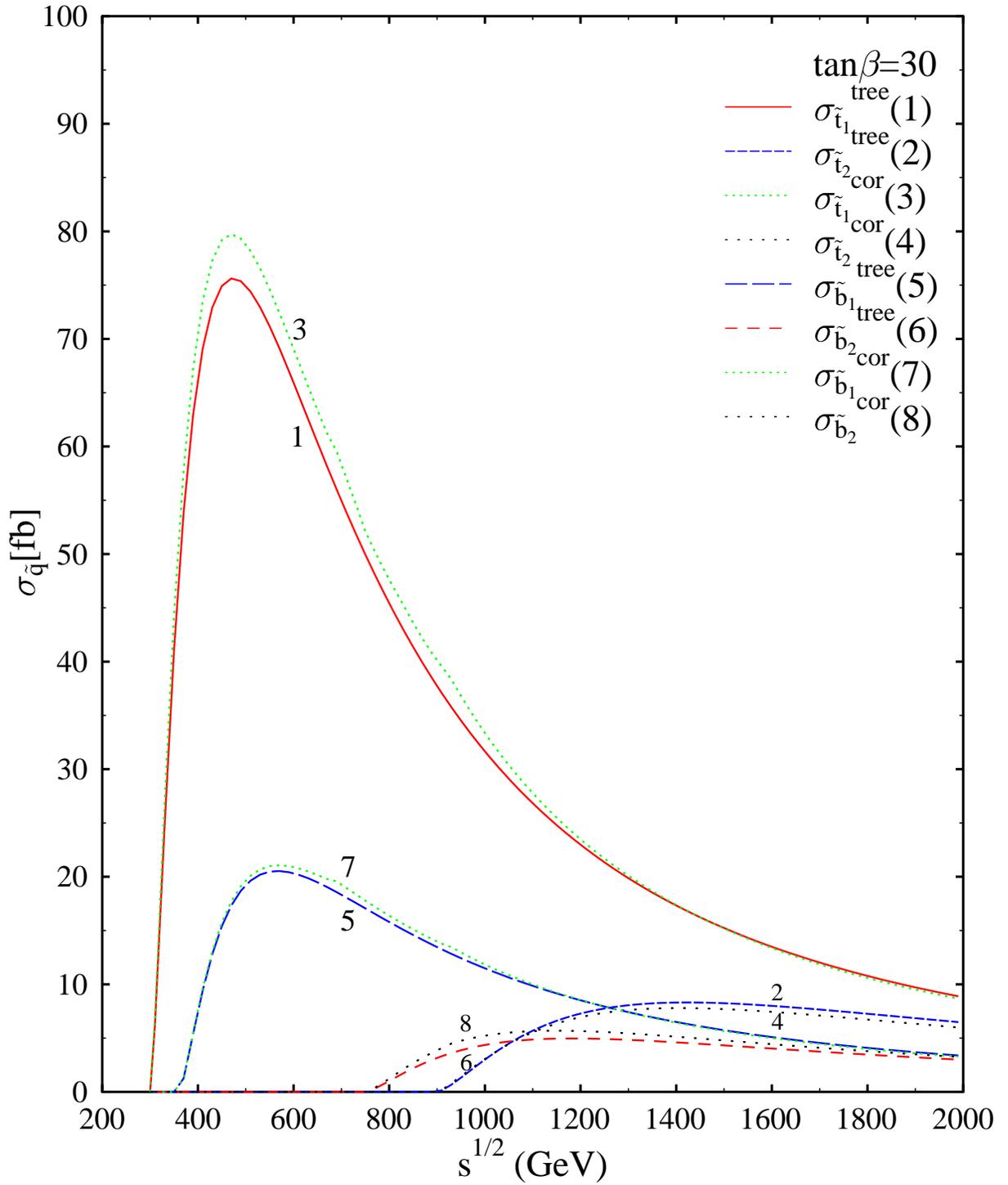
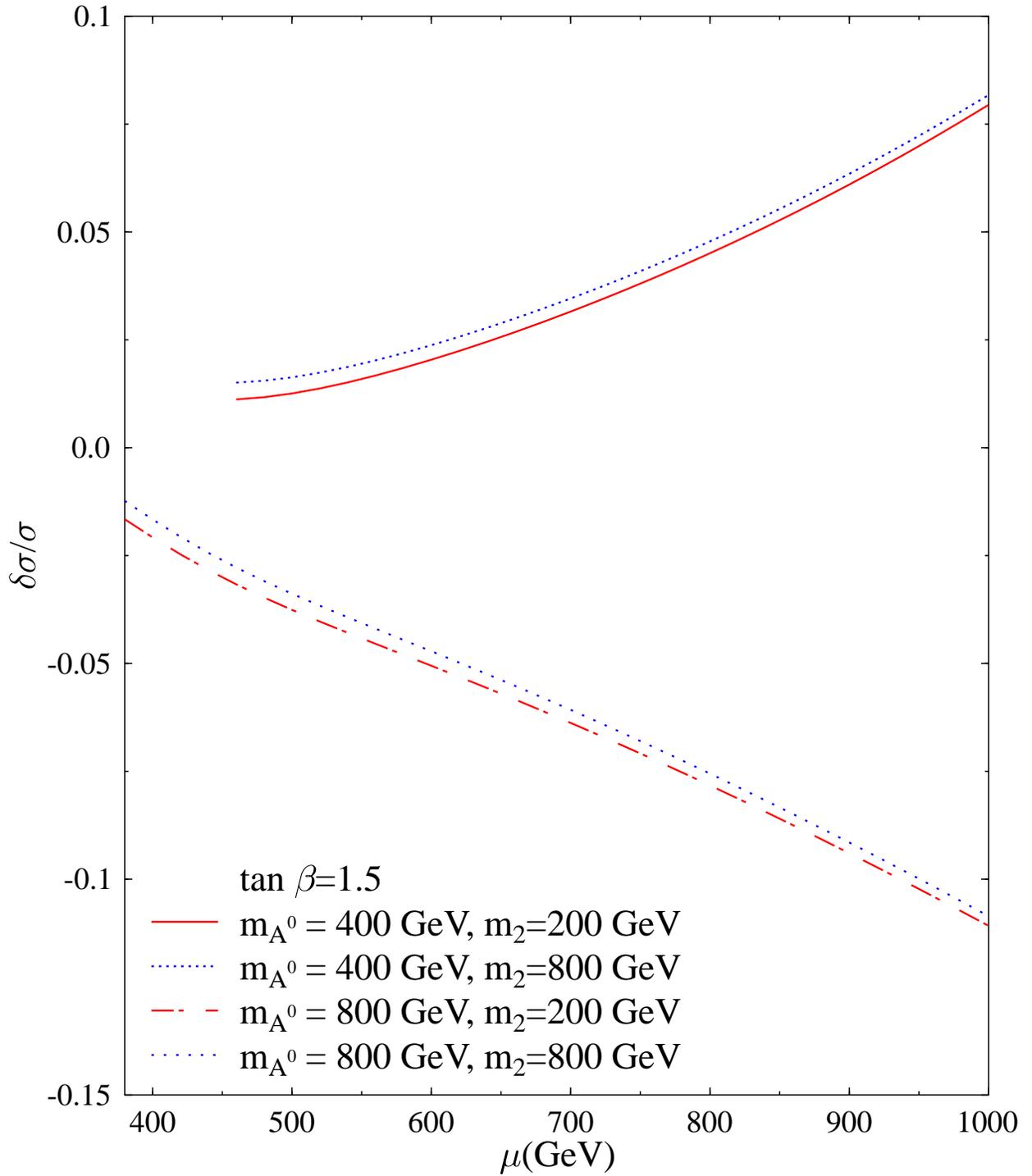
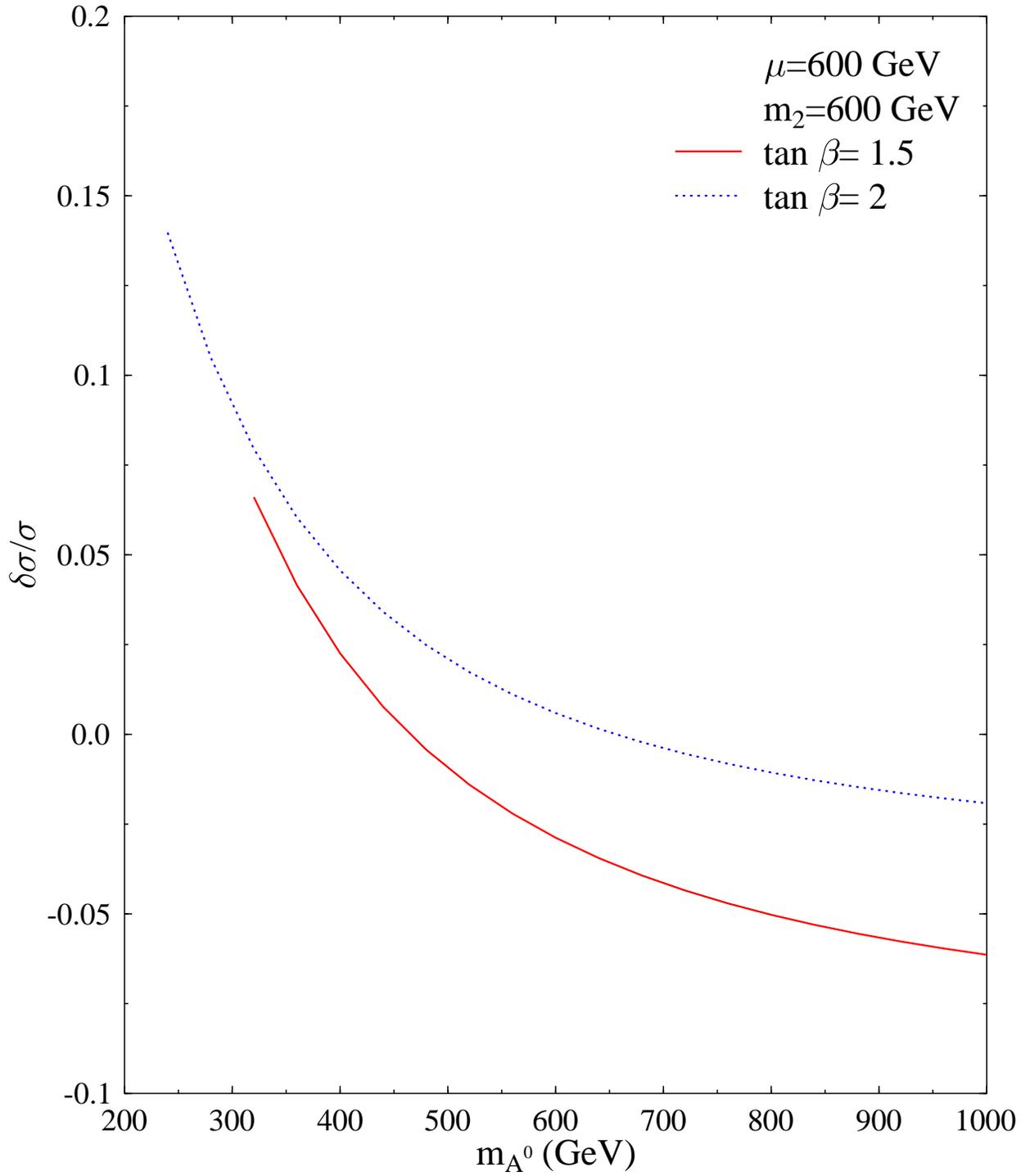
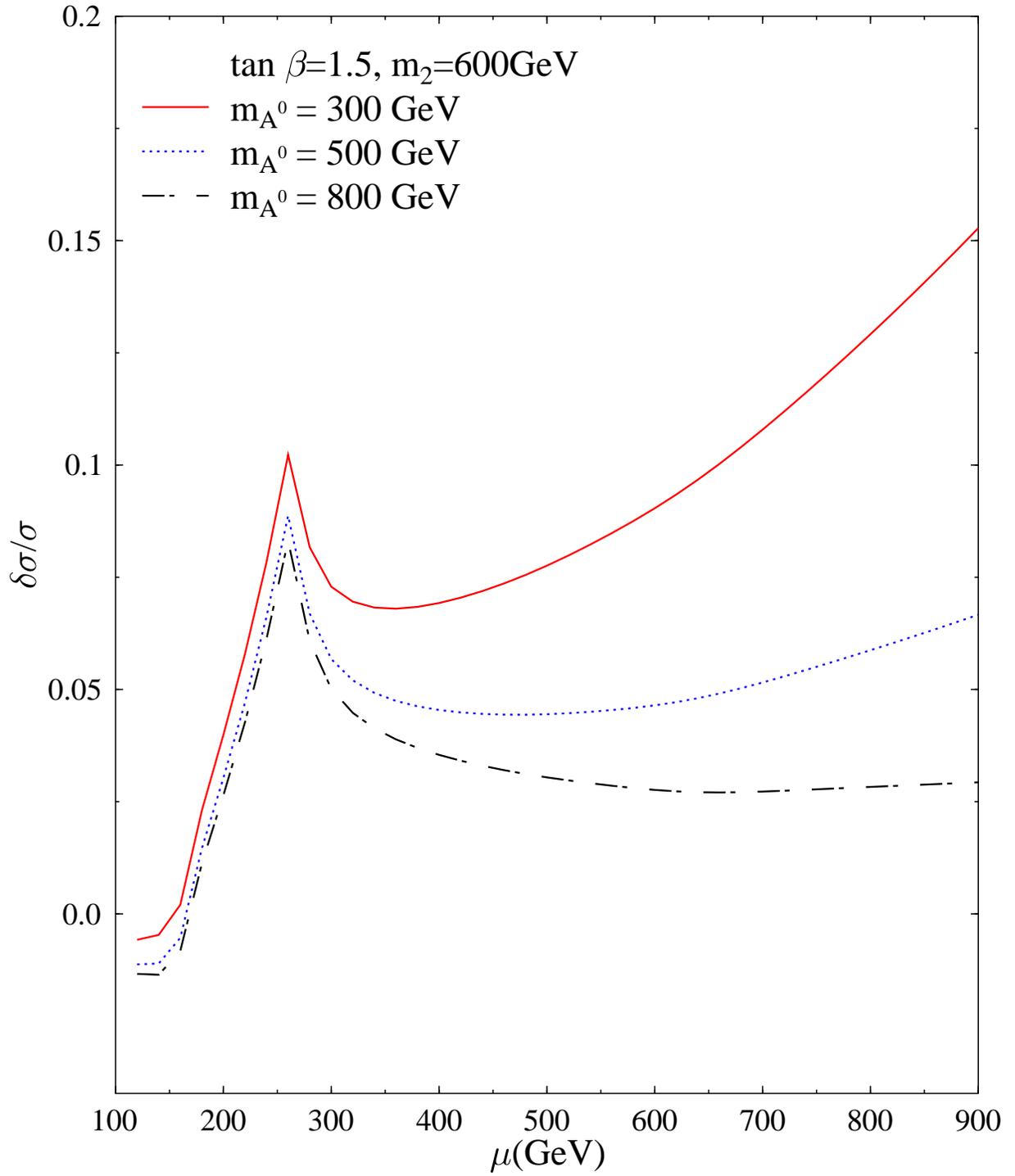
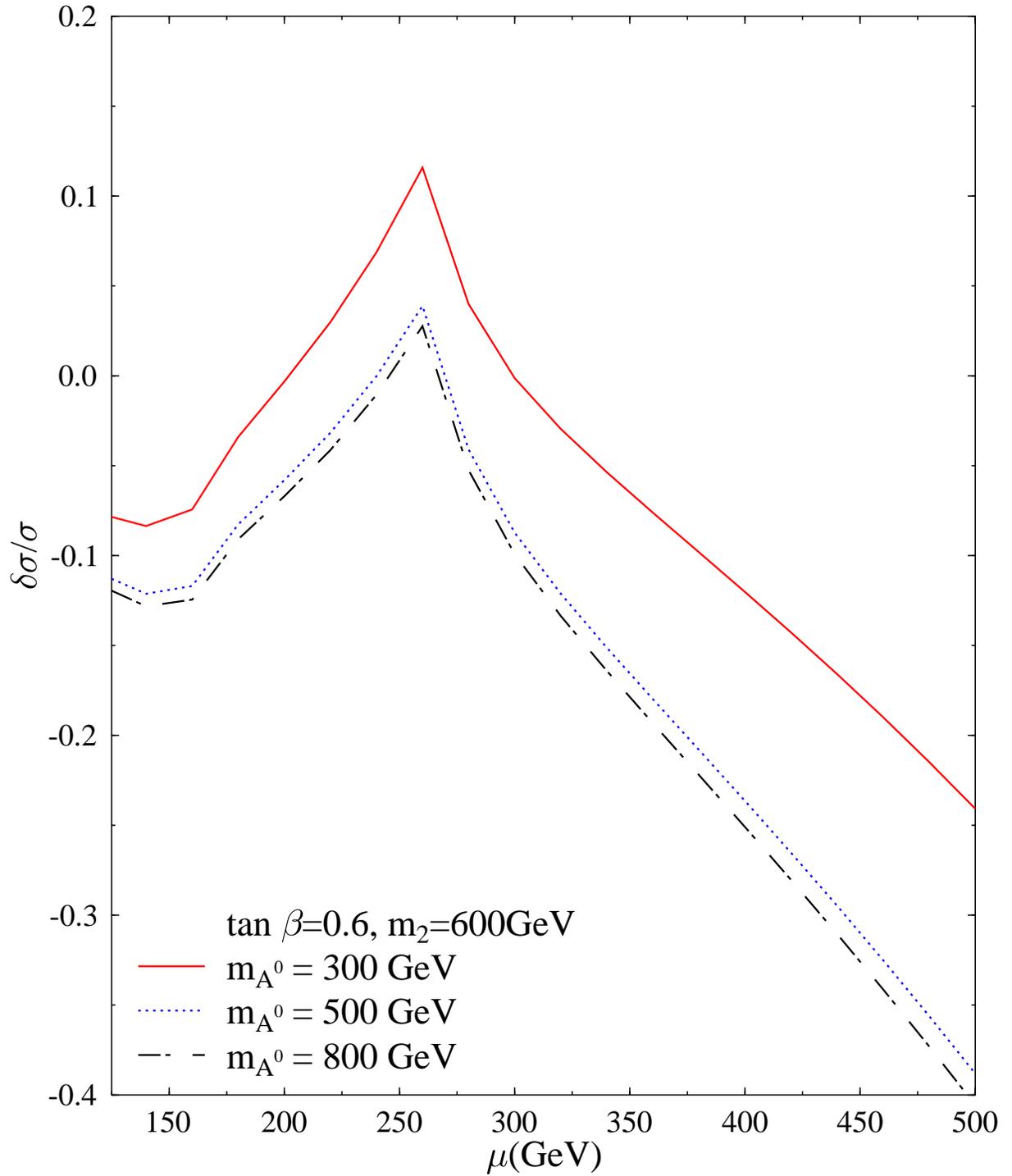
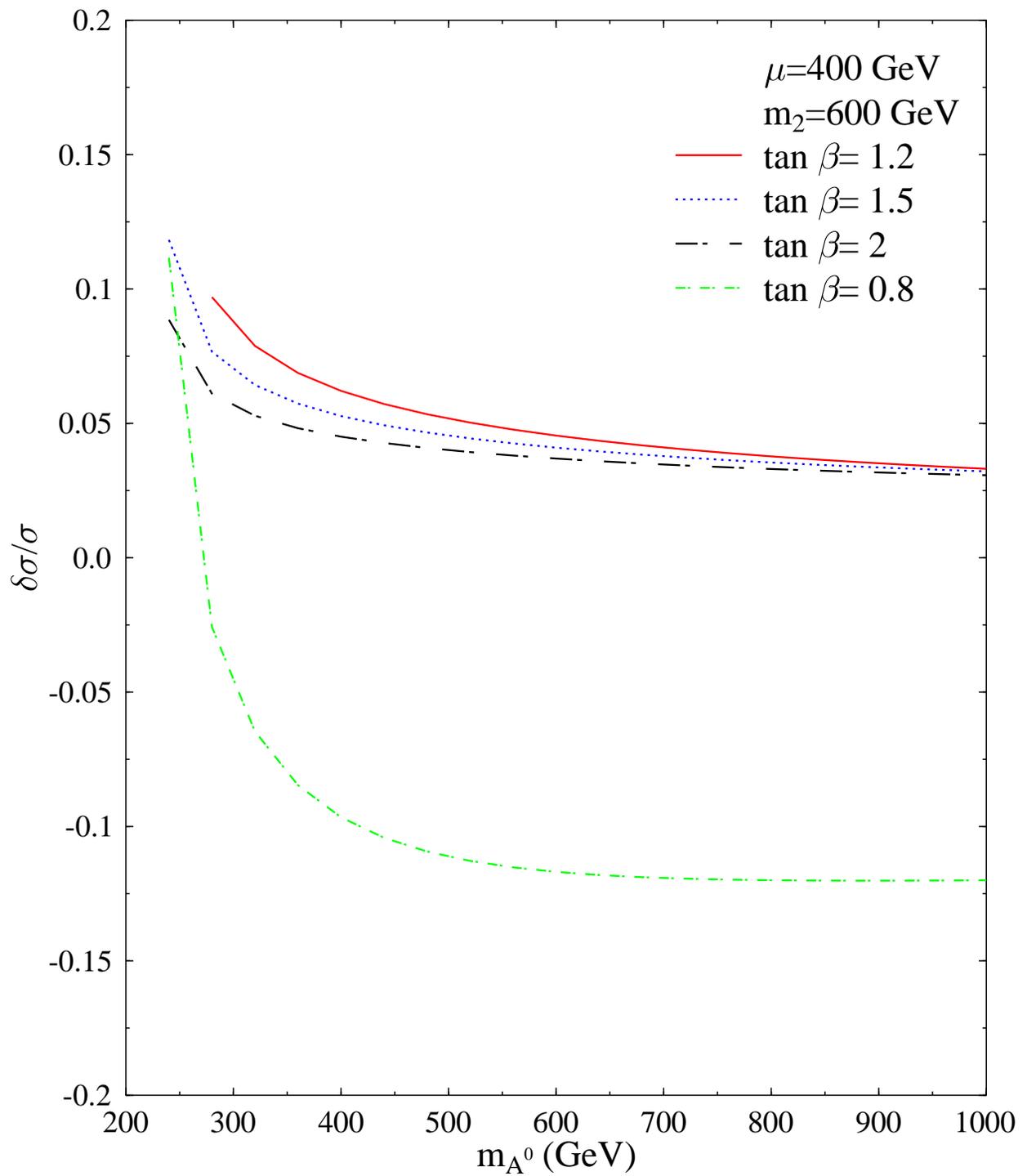
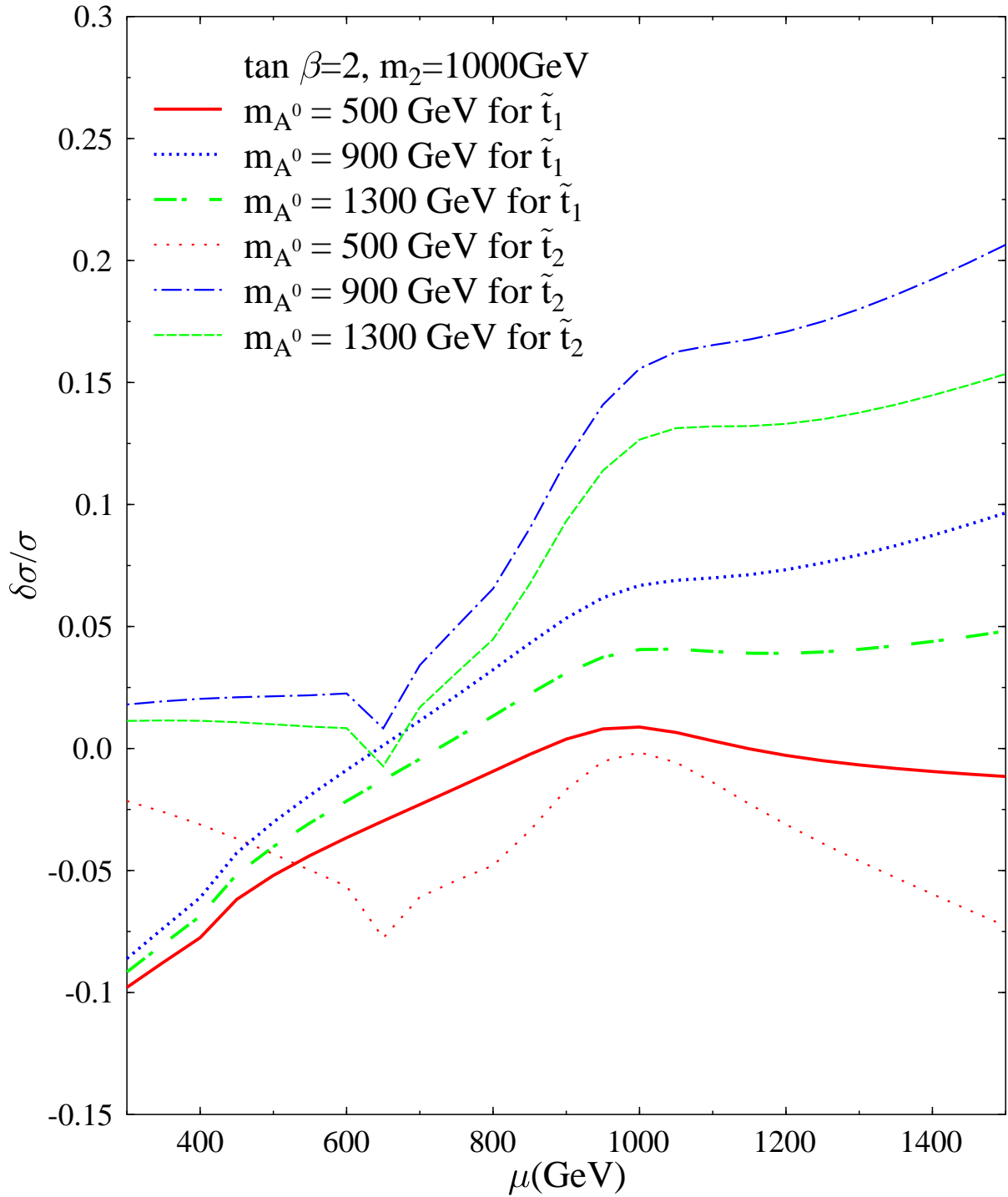
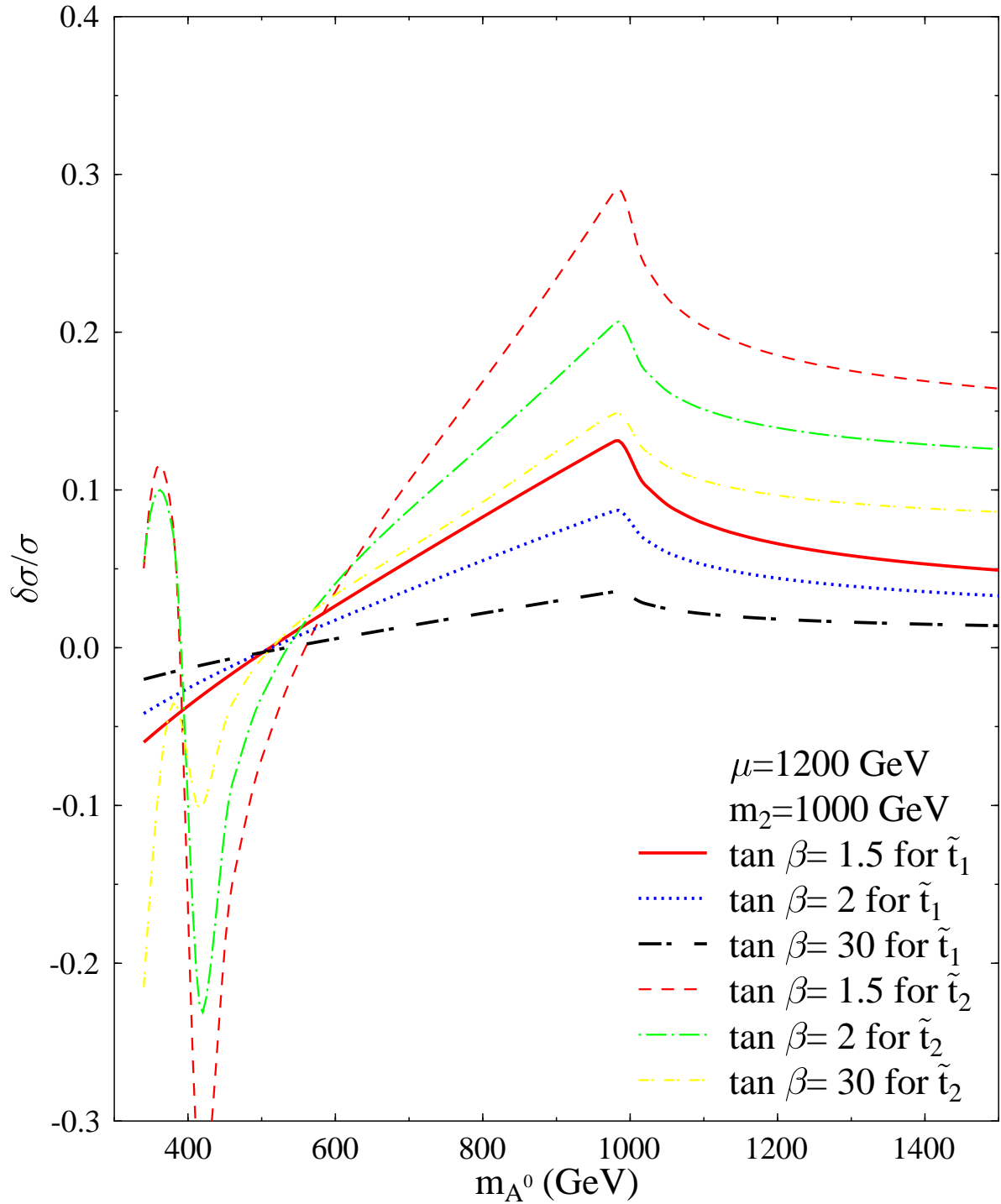
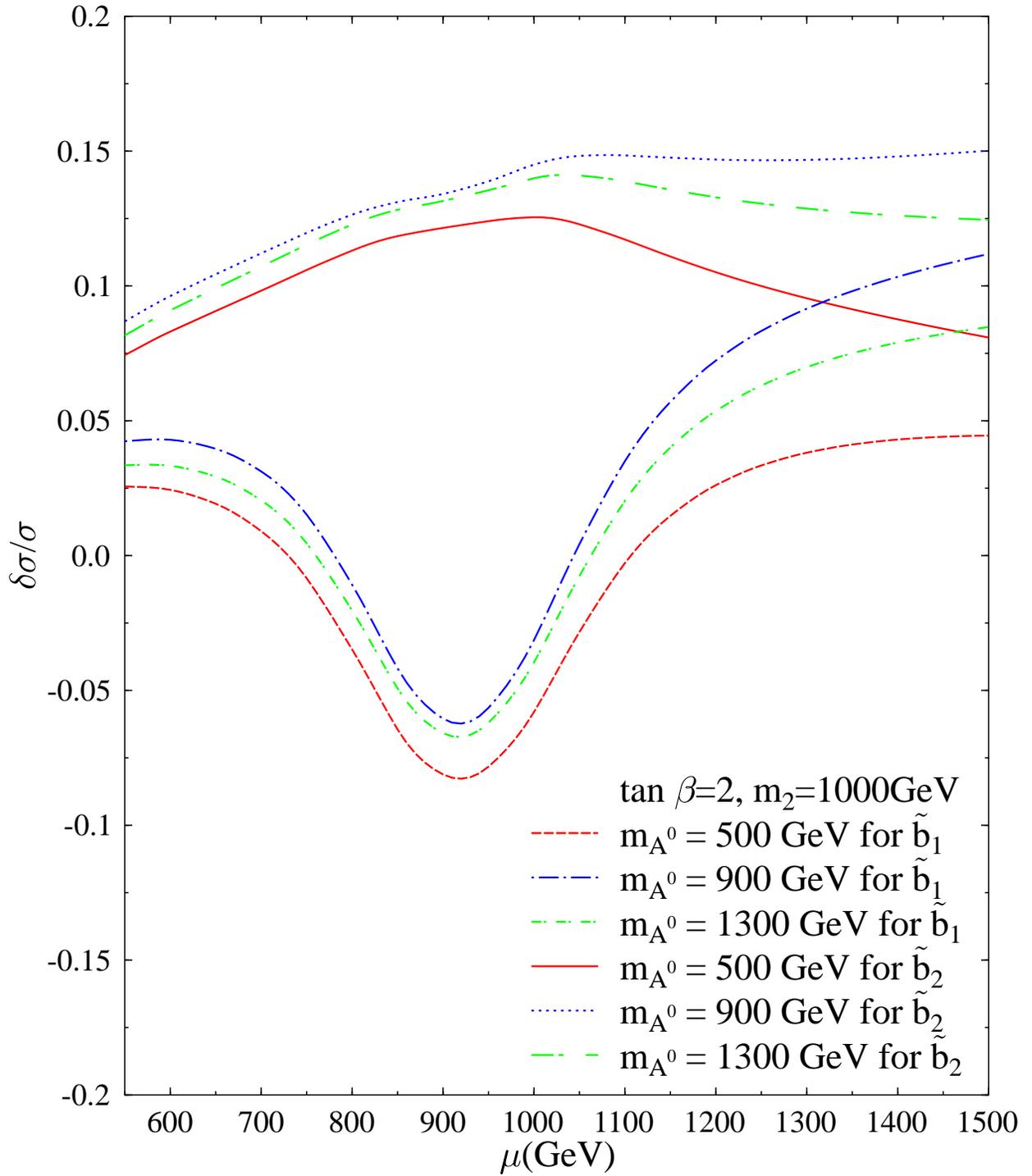
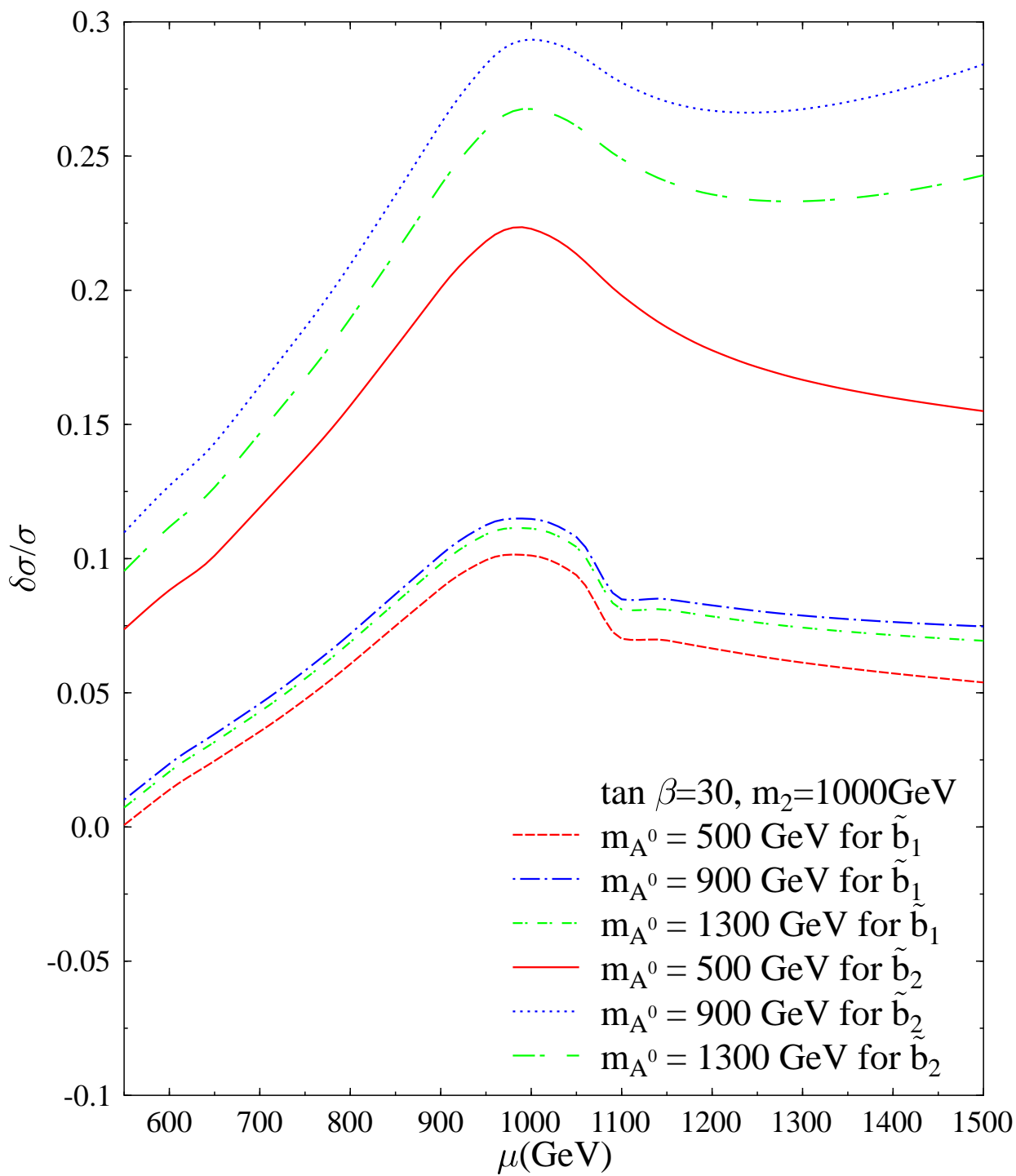
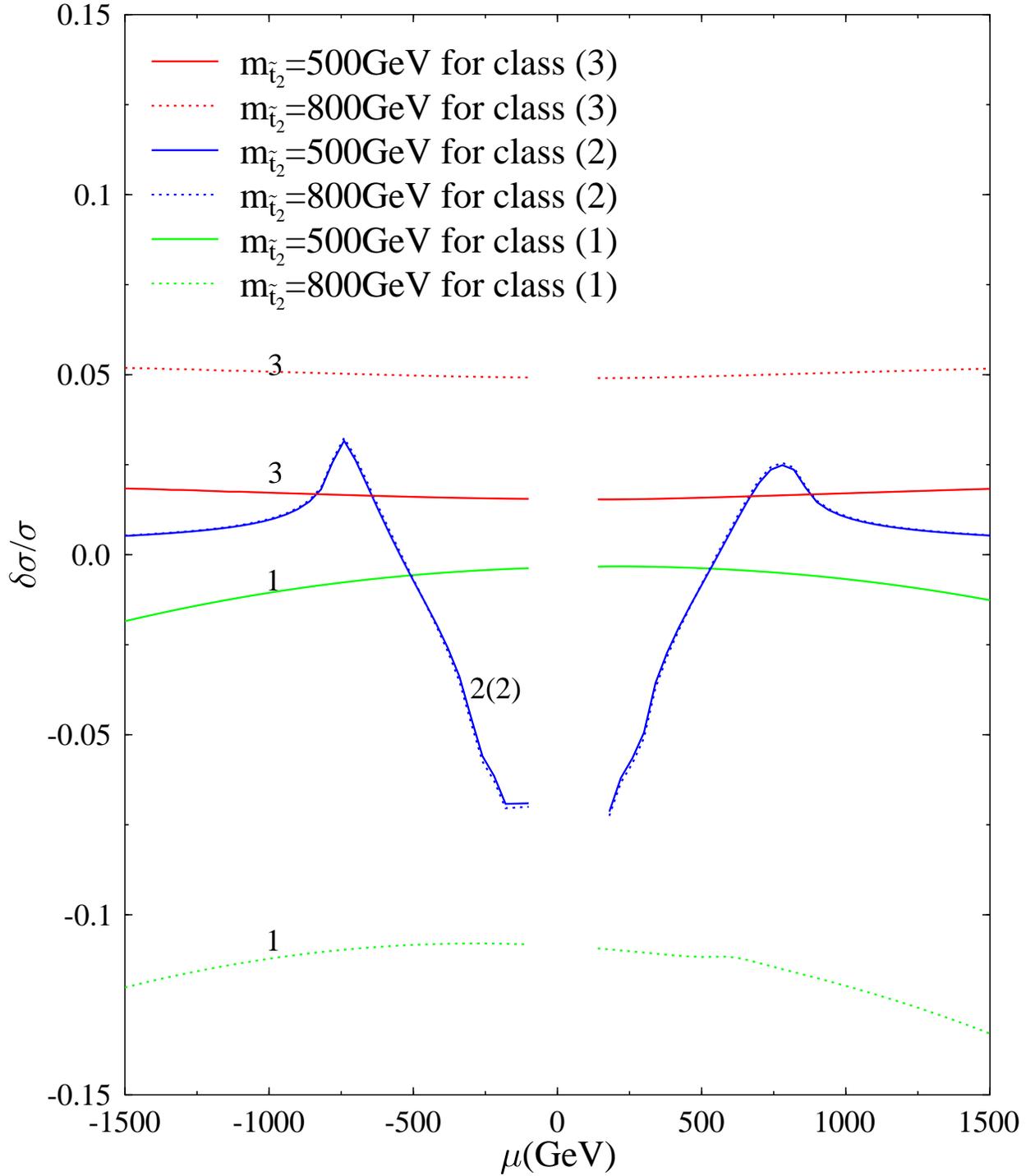
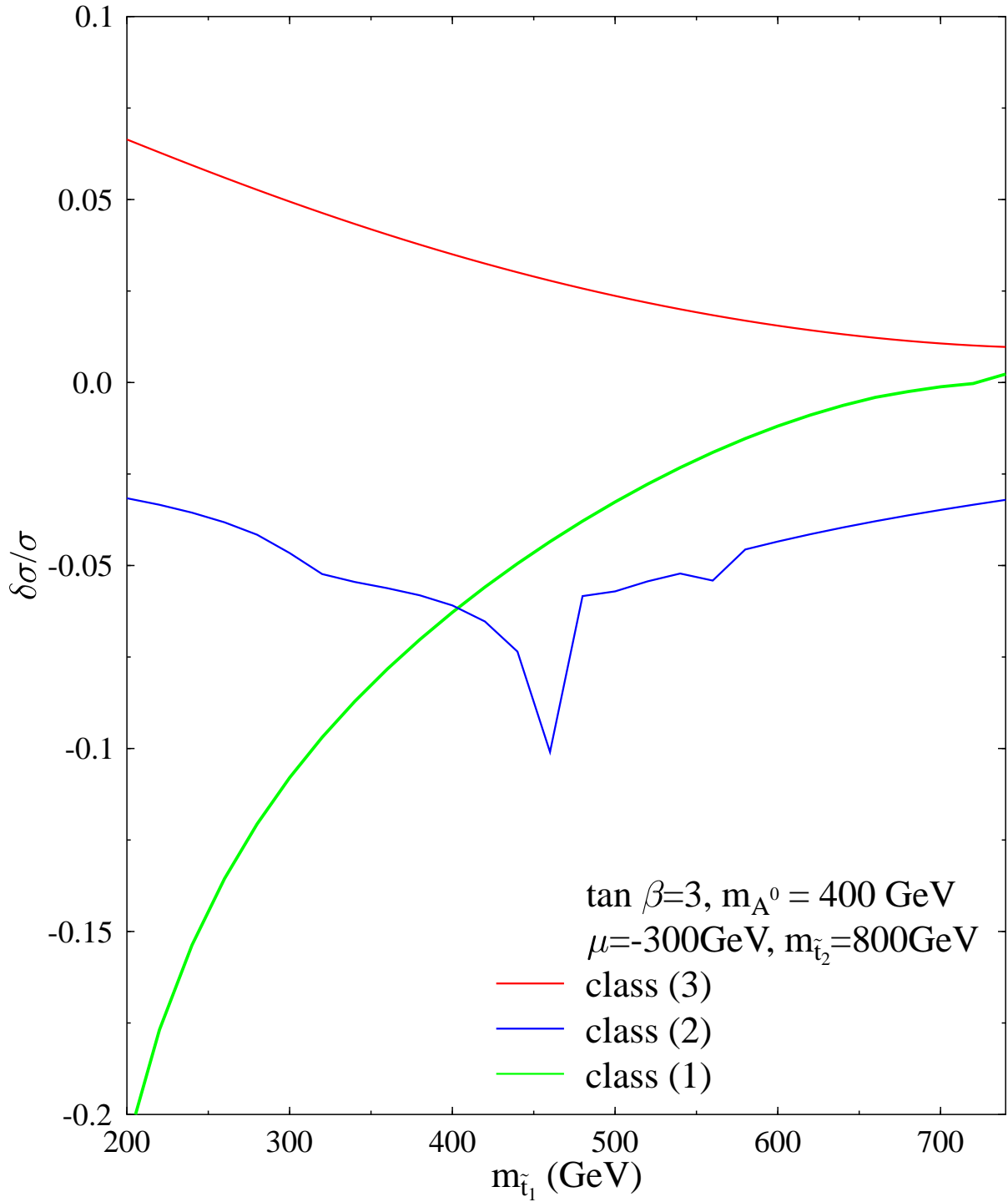
\end{document}